\RequirePackage{fix-cm}
\documentclass[smallextended, natbib]{svjour3}       
\smartqed  
\usepackage{graphicx}
\usepackage{amscd,amsmath,amstext,amsfonts,amsbsy,amssymb}
\usepackage{graphicx,framed,enumerate}
\usepackage{array,multirow,multicol,longtable,booktabs}
\usepackage{threeparttable,rotating,fancyhdr} 

\newcommand{\I}{\mathrm{I}}
\newcommand{\E}{\mathbb{E}}
\newcommand{\V}{\mathbb{V}\mathrm{ar}}

\newcommand{\Pro}{\mathrm{Pr}}

\journalname{Statistical Methods and Applications}
\begin{document}

\title{Estimation of the Volume Under the ROC Surface in Presence of Nonignorable Verification Bias
}

\titlerunning{VUS estimation and NI verification bias}        

\author{Khanh To Duc  \and  Monica Chiogna \and Gianfranco Adimari
}

\institute{Khanh To Duc \\
	Monica Chiogna \\
	Gianfranco Adimari\\
	Department of Statistical Sciences, University of Padua, Via C. Battisti, 241-243, 35121 Padova, Italy.
}

\date{Received: date / Accepted: date}

\maketitle

\begin{abstract}
The volume under the receiver operating characteristic surface (VUS) is useful for measuring the overall accuracy of a diagnostic test when the possible disease status belongs to one of three ordered categories. In medical studies, the VUS of a new test is typically estimated through a sample of measurements obtained by some suitable sample of patients. However, in many cases, only a subset of such patients has the true disease status assessed by a gold standard test. In this paper, for a continuous-scale diagnostic test, we propose four estimators of the VUS which accommodate for nonignorable missingness of the disease status. The estimators are based on a parametric model which jointly describes both the disease and the verification process. Identifiability of the model is discussed. Consistency and asymptotic normality of the proposed estimators are shown, and variance estimation is discussed. The finite-sample behavior is investigated by means of simulation experiments. An illustration is provided.
\end{abstract}

\section{Introduction}
\label{sec:intro}
For an ordinal three-category classification problem, the assessment of the performance of a diagnostic test is achieved by the analysis of the receiver operating characteristic (ROC) surface, which generalizes the ROC curve for binary diagnostic outcomes. The volume under the ROC surface (VUS) is a summary index, usually employed for measuring the overall diagnostic accuracy of the test. Under correct ordering, values of VUS vary from 1/6, suggesting the test is no better than chance alone, to 1, which implies a perfect test, i.e. a test that perfectly discriminates among the three categories. The theoretical construction of the ROC surface and VUS was introduced for the first time by \citet{scu}.

In medical studies, the evaluation of the discriminatory ability of a diagnostic test is typically obtained by making inference about its ROC surface and VUS, based on data from some suitable sample of patients (or units). When the disease status of each patient can be exactly assessed by means of a gold standard (GS) test, a set of methods exist to estimate the ROC surface and VUS of the test in evaluation. See \citet{nak:04}, \citet{xio}, \citet{li} and  \citet{kang}, among others. In practice, however, disease status verification via GS test could be unavailable for all units in the sample, due to the expensiveness and/or invasiveness of the GS test. Thus, often, only a subset of patients undergoes disease verification. In such situations, the implementation of the methods discussed in the above mentioned papers could only be performed on the verified subjects, typically yielding biased estimates of ROC surface and VUS. This bias is known as verification bias.

In order to correct for verification bias, the researchers often assume that the selection for disease verification does not depend on the disease status, given the test results and other observed covariates, i.e., they assume that the true disease status, when missing, is missing at random (MAR, \citet{lit}). Under this assumption, there exist few methods to get bias--corrected inference in ROC surface analysis. \citet{chi} proposed a nonparametric likelihood--based approach to obtain bias--corrected estimators for ROC surface and VUS of an ordinal diagnostic test. In case of continuous diagnostic tests, \citet{toduc} discussed several solutions based on imputation and re--weighting methods, and proposed four verification bias--corrected estimators of the ROC surface and VUS: full imputation (FI), mean score imputation (MSI), inverse probability weighting (IPW) and semi--parametric efficient (SPE) estimators. 

However, in some studies the decision to send a subject to verification may be directly based on the presumed subject's disease status,  or, more generally, the selection mechanism may depend on some unobserved covariates related to disease; in these cases, the MAR assumption does not hold and the missing data mechanism is called nonignorable (NI). 

For two-class problems, methods to deal with NI verification bias have been developed, for instance, in \citet{bak,zho:98,zho:03,zho:04,rot,flu,flu:12,liu}. However, the issue of correcting for NI verification bias in ROC surface analysis is very scarcely considered in the statistical literature. This motivated us to develop bias--corrected methods for continuous diagnostic tests with three--class disease status, under a NI missing data mechanism. In particular, in this paper we adopt parametric regression models for the disease and the verification processes,  extending the selection model of \citet{liu} to match the case of three--class disease status. Then, we use likelihood-based estimators of model parameters to derive four estimators of the VUS. Consistency and asymptotic normality of the proposed estimators are proved. Estimation of their variance is also discussed. 

The rest of the paper is organized as follows. In Section 2, we set the working model and discuss its identifiability. In Section 3 we present our proposed bias-corrected VUS estimators, along with theoretical results about consistency and asymptotic normality. The results of a simulation study are presented in Section 4 and an illustration based on data from the  Alzheimer's Disease Neuroimaging Initiative is provided in Section~5. Concluding remarks are left to Section 6.  Technical details and a discussion about variance estimation can be found in the appendices. 

\section{Model for NI missing data mechanism}
\subsection{Background}
\label{sec:back}
Suppose we need to evaluate the predictive ability of a new continuous diagnostic test in a context where the disease status of a patient can be described by three ordered categories, ``non--diseased'', ``intermediate'' and ``diseased'', say. Consider a sample of $n$ subjects and let $T$, $\vec{\mathcal{D}}$ and $\vec{A}$ denote the test result, the disease status and a vector of covariates for each subject, respectively. In this framework, $\vec{\mathcal{D}}$ can be modeled as a trinomial random vector $\vec{\mathcal{D}} = (D_1, D_2, D_3)^\top$, such that $D_k$ is a Bernoulli random variable having mean $\theta_k = \Pro(D_k = 1)$ where $\theta_1 + \theta_2 + \theta_3 = 1$. Hence,  $\theta_k$ represents the probability that a generic subject, classified according to its disease status, belong to the class $k$. We are interested in estimating the VUS of the test $T$, say $\mu$, which is defined as
(\citet{nak:04})
\begin{eqnarray}
\mu &=& \Pro \left(T_i < T_\ell < T_r|D_{1i} = 1,D_{2\ell} = 1,D_{3r} = 1\right) \nonumber \\
&& + \: \frac{1}{2} \Pro\left(T_i < T_\ell = T_r|D_{1i} = 1,D_{2\ell} = 1,D_{3r} = 1\right) \nonumber\\
&& + \: \frac{1}{2} \Pro\left(T_i = T_\ell < T_r|D_{1i} = 1,D_{2\ell} = 1,D_{3r} = 1\right) \nonumber\\
&& + \: \frac{1}{6} \Pro\left(T_i = T_\ell = T_r|D_{1i} = 1,D_{2\ell} = 1,D_{3r} = 1\right) \nonumber
\end{eqnarray}
or, equivalently,
\begin{equation}
\mu = \frac{\E \left(D_{1i}D_{2\ell}D_{3r}\I_{i\ell r}\right)}{\E \left(D_{1i}D_{2\ell}D_{3r}\right)}, \label{org:vus}
\end{equation}
where the indices $i$, $\ell$, $r$ refer to three different subjects, $\I_{i\ell r} = \I(T_i < T_\ell < T_r) + 1/2 \I(T_i < T_\ell = T_r) + 1/2 \I(T_i = T_\ell < T_r) + 1/6 \I(T_i = T_\ell = T_r)$ and $\I(\cdot)$ is the indicator function.

When the disease status $\vec{\mathcal{D}}$ is available for all subjects, a natural nonparametric estimator of $\mu$ is  given by
\begin{equation}
\hat{\mu}_{\mathrm{NP}} = \frac{\sum\limits_{i=1}^{n}\sum\limits_{\ell = 1, \ell \ne i}^{n} \sum\limits_{\stackrel{r = 1}{r \ne \ell, r \ne i}}^{n}\I_{i\ell r}D_{1i}D_{2\ell}D_{3r}}{\sum\limits_{i=1}^{n}\sum\limits_{\ell = 1, \ell \ne i}^{n} \sum\limits_{\stackrel{r = 1}{r \ne \ell, r \ne i}}^{n} D_{1i}D_{2\ell}D_{3r}}.
\label{nonp:vus}
\end{equation}
However, in many situations not all subjects undergo the verification process, and hence, the disease status $\vec{\mathcal{D}}$ is missing in a subset of patients in the study. Let $V_i$ be the verification status for the $i$-th subject: $V_i = 1$ if $\vec{\mathcal{D}}_i$ is observed and $V_i = 0$ otherwise. We define the observed data as the set $\{O_i = (\vec{\mathcal{D}}_i^\top, V_i, T_i, \vec{A}^\top_i)^\top$, \ $i = 1,\ldots,n\}$. When the true disease status is subject to NI missingness, estimators working under the MAR assumption cannot be applied tout court. Our goal is to adjust FI, MSI, IPW and SPE estimators discussed in \cite{toduc} to the framework of NI missingness.

\subsection{Model settings}
\label{sec:mdset}
To deal with NI missing data mechanism, in what follows we extend parametric models adopted in \citet{liu} for the two--class problem to the three--class case. More precisely, with three disease categories, we fix the model for the verification process as follows
\begin{equation}
\pi = \Pro(V = 1| D_1, D_2, T, \vec{A}) = \frac{\exp\left\{h(T, \vec{A}; \vec{\tau}_\pi) + \lambda_1 D_1 + \lambda_2 D_2 \right\}}{1 + \exp\left\{h(T, \vec{A}; \vec{\tau}_\pi) + \lambda_1 D_1 + \lambda_2 D_2 \right\}},
\label{veri:model:2}
\end{equation}
where $D_1$ and $D_2$ are defined in the previous section, $h(T, \vec{A}; \vec{\tau}_\pi)$ is, in general,  an arbitrary working function, and $\vec{\tau}_\pi$ is a set of parameters. Here, $\vec{\lambda} = (\lambda_1,\lambda_2)^\top$ is the non-ignorable parameter: the missing data mechanism is MAR if $\lambda_1 = \lambda_2 = 0$; NI, otherwise. As for the disease model, we employ the multinomial logistic regression for the whole sample, i.e.,
\begin{eqnarray}
\rho_k = \Pro(D_k = 1| T, \vec{A}) = \frac{\exp\left\{f(T, \vec{A}; \vec{\tau}_{\rho_k})\right\}}{1 + \exp\left\{f(T, \vec{A}; \vec{\tau}_{\rho_1})\right\} + \exp\left\{f(T, \vec{A}; \vec{\tau}_{\rho_2})\right\}},
\label{dise:model}
\end{eqnarray}
where $f(T, \vec{A}; \vec{\tau}_{\rho_k})$ is an arbitrary working function, and $\vec{\tau}_{\rho_k}$ is a set of parameters, for $k = 1,2$. The parameters $\vec{\lambda}, \vec{\tau}_\pi, \vec{\tau}_\rho$, with $\vec{\tau}_{\rho} = (\vec{\tau}^\top_{\rho_1}, \vec{\tau}^\top_{\rho_2})^\top$, can be estimated jointly by using a likelihood--based approach.

It is worth noting that, under (\ref{veri:model:2}), an application of Bayes' rule gives that
\begin{eqnarray}
\frac{\Pro(D_1 = 1|V = 1, T, \vec{A})}{\Pro(D_1 = 1|V = 0, T, \vec{A})} &=& \frac{\Pro(V = 0| T, \vec{A})}{\Pro(V = 1|T, \vec{A})} \exp\left\{h(T, \vec{A}; \vec{\tau}_\pi) + \lambda_1\right\} \nonumber ,\\
\frac{\Pro(D_2 = 1|V = 1, T, \vec{A})}{\Pro(D_2 = 1|V = 0, T, \vec{A})} &=& \frac{\Pro(V = 0|T, \vec{A})}{\Pro(V = 1|T, \vec{A})} \exp\left\{h(T, \vec{A}; \vec{\tau}_\pi) + \lambda_2\right\} \nonumber ,\\
\frac{\Pro(D_3 = 1|V = 1, T, \vec{A})}{\Pro(D_3 = 1|V = 0, T, \vec{A})} &=& \frac{\Pro(V = 0|T,A)}{\Pro(V = 1|T, \vec{A})} \exp\left\{h(T, \vec{A}; \vec{\tau}_\pi)\right\} \nonumber.
\end{eqnarray}
Therefore,
\begin{eqnarray}
\frac{\Pro(D_1 = 1|V = 1, T, \vec{A})}{\Pro(D_1 = 1|V = 0, T, \vec{A})} \bigg / \frac{\Pro(D_3 = 1|V = 1, T, \vec{A})}{\Pro(D_3 = 1|V = 0, T, \vec{A})} &=& \exp(\lambda_1) , \label{inter:lambda_1} \\
\frac{\Pro(D_2 = 1|V = 1, T, \vec{A})}{\Pro(D_2 = 1|V = 0, T, \vec{A})} \bigg / \frac{\Pro(D_3 = 1|V = 1, T, \vec{A})}{\Pro(D_3 = 1|V = 0, T, \vec{A})} &=& \exp(\lambda_2) , \label{inter:lambda_2}
\end{eqnarray}
so that, according to (\ref{inter:lambda_1}) and (\ref{inter:lambda_2}), $\lambda_1$ and $\lambda_2$ can also be interpreted as log-odds ratios of belonging to class 1 (instead of class 3) and to class 2 (instead of class 3), respectively, for a verified subject compared to an unverified subject with the same test result and covariates. 

\subsection{Parameter estimation}
\label{sec:para_est}
As in \citet{liu}, in our model, for simplicity, we take $h(T, \vec{A}; \vec{\tau}_\pi) = \tau_{\pi_1} + \tau_{\pi_2} T + \vec{A}^\top \vec{\tau}_{\pi_3}$ and  $f(T, \vec{A}; \vec{\tau}_{\rho_k}) = \tau_{\rho_{1k}} + \tau_{\rho_{2k}} T + \vec{A}^\top \vec{\tau}_{\rho_{3k}}$, which is a natural choice in practice. For fixed $T$ and $\vec{A}$, the observed distribution is fully determined by the three probabilities $\Pro(D_1 = 1, D_2 = 0, V = 1| T, \vec{A})$, $\Pro(D_1 = 0, D_2 = 1, V = 1| T, \vec{A})$ and $\Pro(D_1 = 0, D_2 = 0, V = 1| T, \vec{A})$. It is easy to show that
\begin{eqnarray}
\lefteqn{\Pro(D_1 = 1, D_2 = 0, V = 1| T, \vec{A})} \nonumber \\
&=& \Pro(D_1 = 1, D_2 = 0| T, \vec{A})\Pro(V = 1| D_1 = 1, D_2 = 0, T, \vec{A}) \nonumber \\
&=& \Pro(D_1 = 1| T, \vec{A}) \Pro(V = 1| D_1 = 1, D_2 = 0, T, \vec{A}) \nonumber \\
&=& \rho_{1} \pi_{10} \nonumber.
\end{eqnarray}
Similarly, we have that
\begin{eqnarray}
\Pro(D_1 = 0, D_2 = 1, V = 1| T, \vec{A}) &=& \rho_{2}\pi_{01} \nonumber, \\
\Pro(D_1 = 0, D_2 = 0, V = 1| T, \vec{A}) &=& (1 - \rho_{1} - \rho_{2})\pi_{00} \nonumber,
\end{eqnarray}
with 
$\pi_{01}=\Pro(V = 1| D_1 = 0, D_2 = 1, T, \vec{A}) $ and  
$\pi_{00}=\Pro(V = 1| D_1 = 0, D_2 = 0, T, \vec{A}) $.
Then,
\begin{eqnarray}
\Pro(V = 1|T, \vec{A}) =  \rho_1 \pi_{10} + \rho_2 \pi_{01} + (1 - \rho_1 - \rho_2)\pi_{00} , \nonumber
\end{eqnarray}
and  $\Pro(V = 0|T, \vec{A}) = 1 - \Pro(V = 1|T, \vec{A}) = 1 - \rho_1 \pi_{10} + \rho_2 \pi_{01} + (1 - \rho_1 - \rho_2)\pi_{00}$. It follows that the log-likelihood function can be written as
\begin{eqnarray}
\lefteqn{\log L (\vec{\lambda}, \vec{\tau}_\pi, \vec{\tau}_\rho)} \nonumber\\
&=& \sum_{i = 1}^{n}\bigg\{ D_{1i} V_i \log(\rho_{1i}\pi_{10i}) + D_{2i}V_i \log(\rho_{2i}\pi_{01i}) \nonumber\\
&& + \: (1 - D_{1i} - D_{2i})V_i \log((1 - \rho_{1i} - \rho_{2i})\pi_{00i}) \nonumber\\
&& + \: (1 - V_i) \log(1 - \rho_{1i} \pi_{10i} - \rho_{2i} \pi_{01i} - (1 - \rho_{1i} - \rho_{2i})\pi_{00i}) \bigg\} .
\label{lg-like}
\end{eqnarray}
The estimates $\hat{\vec{\lambda}}$, $\hat{\vec{\tau}}_\pi$, and $\hat{\vec{\tau}}_\rho$ can be obtained by maximizing $\log L(\vec{\lambda}, \vec{\tau}_\pi, \vec{\tau}_\rho)$ or by solving the score equations
\begin{eqnarray}
0 &=& \sum_{i = 1}^{n}\left\{D_{1i}V_i(1 - \pi_{10i}) - \frac{(1 - V_i) \rho_{1i}\pi_{10i} (1 - \pi_{10i}) }{1 - \rho_{1i} \pi_{10i} - \rho_{2i} \pi_{01i} - (1 - \rho_{1i} - \rho_{2i})\pi_{00i}} \right\} \nonumber, \\
0 &=& \sum_{i = 1}^{n}\left\{D_{2i}V_i(1 - \pi_{01i}) - \frac{(1 - V_i) \rho_{2i}\pi_{01i} (1 - \pi_{01i})}{1 - \rho_{1i} \pi_{10i} - \rho_{2i} \pi_{01i} - (1 - \rho_{1i} - \rho_{2i})\pi_{00i}} \right\} \nonumber, \\
0 &=& \sum_{i = 1}^{n}\vec{U}_i \bigg\{ D_{1i}V_i(1 - \pi_{10i}) + D_{2i}V_i(1 - \pi_{01i}) + (1 - D_{1i} - D_{2i})V_i(1 - \pi_{00i}) \nonumber \\
&& - \: (1 - V_i)\frac{\rho_{1i}\pi_{10i}(1 - \pi_{10i}) + \rho_{2i}\pi_{01i}(1 - \pi_{01i}) + (1 - \rho_{1i} - \rho_{2i})\pi_{00i}(1 - \pi_{00i})}{1 - \rho_{1i} \pi_{10i} - \rho_{2i} \pi_{01i} - (1 - \rho_{1i} - \rho_{2i})\pi_{00i}} \bigg\} \nonumber, \\
0 &=& \sum_{i = 1}^{n}\vec{U}_i \left\{ V_i(D_{1i} - \rho_{1i}) - (1 - V_i)\frac{(\pi_{10i} - \pi_{00i})\rho_{1i}(1 - \rho_{1i}) - (\pi_{01i} - \pi_{00i})\rho_{1i}\rho_{2i}}{1 - \rho_{1i} \pi_{10i} - \rho_{2i} \pi_{01i} - (1 - \rho_{1i} - \rho_{2i})\pi_{00i}} \right\} \nonumber, \\
0 &=& \sum_{i = 1}^{n}\vec{U}_i \left\{ V_i(D_{2i} - \rho_{2i}) - (1 - V_i)\frac{(\pi_{01i} - \pi_{00i})\rho_{2i}(1 - \rho_{2i}) - (\pi_{10i} - \pi_{00i})\rho_{1i}\rho_{2i}}{1 - \rho_{1i} \pi_{10i} - \rho_{2i} \pi_{01i} - (1 - \rho_{1i} - \rho_{2i})\pi_{00i}} \right\} \nonumber,
\end{eqnarray}
where $\vec{U}_i = (1, T_i, \vec{A}^\top_i)^\top$. The above equations are obtained by using the following results
\begin{eqnarray}
\frac{\partial}{\partial \lambda_1}\pi_{10i} &=& \pi_{10i}(1 - \pi_{10i}), \nonumber \\
\frac{\partial}{\partial \lambda_2}\pi_{01i} &=& \pi_{01i}(1 - \pi_{01i}), \nonumber \\
\frac{\partial}{\partial \vec{\tau}_\pi^\top}\pi_{d_1 d_2 i} &=& \vec{U}_i (1 - \pi_{d_1 d_2 i})\pi_{d_1 d_2 i} \nonumber
\end{eqnarray}
(here $(d_1, d_2)$ is a pair in the set $\{(1,0), (0,1), (0,0)\}$), and
\begin{equation}
\begin{array}{r l r l}
\dfrac{\partial}{\partial \vec{\tau}^\top_{\rho_1}} \rho_{1i} &= \vec{U}_i\rho_{1i}(1 - \rho_{1i}); & \qquad  \dfrac{\partial}{\partial \vec{\tau}^\top_{\rho_2}} \rho_{1i} &= - \vec{U}_i\rho_{1i}\rho_{2i}; \\ [16pt]
\dfrac{\partial}{\partial \vec{\tau}^\top_{\rho_2}} \rho_{2i} &= \vec{U}_i\rho_{2i}(1 - \rho_{2i}); & \qquad \dfrac{\partial}{\partial \vec{\tau}^\top_{\rho_1}} \rho_{2i} &= - \vec{U}_i \rho_{1i}\rho_{2i}.
\end{array}
\nonumber
\end{equation}
\subsection{Identifiability}
\label{sec:ident}
In this section, we verify that the working model  based on 
(\ref{veri:model:2}), with  $h(T, \vec{A}; \vec{\tau}_\pi) = \tau_{\pi_1} + \tau_{\pi_2} T + \vec{A}^\top \vec{\tau}_{\pi_3}$, and (\ref{dise:model}), with $f(T, \vec{A}; \vec{\tau}_{\rho_k}) = \tau_{\rho_{1k}} + \tau_{\rho_{2k}} T + \vec{A}^\top \vec{\tau}_{\rho_{3k}}$, is identifiable. Since the log--likelihood (\ref{lg-like}) is fully determined by the three probabilities $\Pro(D_1 = 1, D_2 = 0, V = 1| T, \vec{A})$, $\Pro(D_1 = 0, D_2 = 1, V = 1| T, \vec{A})$ and $\Pro(D_1 = 0, D_2 = 0, V = 1| T, \vec{A})$, we have to show that such probabilities are uniquely determined by the parameters for all possible $T$ and $\vec{A}$. For the sake of simplicity, in the remainder of this section the auxiliary covariate $\vec{A}$ is omitted (actually, we can always view $\vec{A}$ as fixed while varying $T$). 

Let $\vec{\xi} = (\lambda_1, \lambda_2, \tau_{\pi_1}, \tau_{\pi_2}, \tau_{\rho_{11}}, \tau_{\rho_{21}}, \tau_{\rho_{12}}, \tau_{\rho_{22}})^\top$ be the set of parameters. For given $T = t$, we can write
\begin{eqnarray}
\log(\rho_{1}\pi_{10}) &=& (\tau_{\rho_{11}} + \tau_{\rho_{21}}t) - \log\left\{1 + \exp(\tau_{\rho_{11}} + \tau_{\rho_{21}}t) + \exp(\tau_{\rho_{12}} + \tau_{\rho_{22}}t) \right\} \nonumber \\
&& + \: (\tau_{\pi_1} + \tau_{\pi_2} t) + \lambda_1 - \log \left\{1 + \exp(\tau_{\pi_1} + \tau_{\pi_2} t)\exp(\lambda_1) \right\} \nonumber, \\
\log(\rho_{2}\pi_{01}) &=& (\tau_{\rho_{12}} + \tau_{\rho_{22}}t) - \log\left\{1 + \exp(\tau_{\rho_{11}} + \tau_{\rho_{21}}t) + \exp(\tau_{\rho_{12}} + \tau_{\rho_{22}}t) \right\} \nonumber \\
&& + \: (\tau_{\pi_1} + \tau_{\pi_2} t) + \lambda_2 - \log \left\{1 + \exp(\tau_{\pi_1} + \tau_{\pi_2} t)\exp(\lambda_2) \right\} \nonumber, \\
\log(\rho_{3} \pi_{00}) &=& - \log\left\{1 + \exp(\tau_{\rho_{11}} + \tau_{\rho_{21}}t) + \exp(\tau_{\rho_{12}} + \tau_{\rho_{22}}t) \right\} + (\tau_{\pi_1} + \tau_{\pi_2} t) \nonumber \\
&& - \: \log \left\{1 + \exp(\tau_{\pi_1} + \tau_{\pi_2} t) \right\} \nonumber.
\end{eqnarray}
Let $x(t) = \tau_{\pi_1} + \tau_{\pi_2} t$, $y(t) = \tau_{\rho_{11}} + \tau_{\rho_{21}}t$ and $z(t) = \tau_{\rho_{12}} + \tau_{\rho_{22}}t$,  for each $t \in \mathbb{R}$. The above expressions, which refer to the quantities characterizing the  log--likelihood function (\ref{lg-like}), can be rewritten as
\begin{eqnarray}
\log(\rho_{3} \pi_{00}) &=& - \log\left\{1 + \exp(y(t)) + \exp(z(t)) \right\} + x(t) - \log \left\{1 + \exp(x(t)) \right\}, \nonumber \\
\log(\rho_{1} \pi_{10}) &=& y(t) - \log\left\{1 + \exp(y(t)) + \exp(z(t)) \right\} + x(t) + \lambda_1 \nonumber\\
&& - \: \log \left\{1 + \exp(x(t))\exp(\lambda_1) \right\} \nonumber \\
&=& \log(\rho_{3} \pi_{00}) + \log \left\{1 + \exp(x(t)) \right\} + y(t) + \lambda_1 \nonumber\\
&& - \: \log \left\{1 + \exp(x(t))\exp(\lambda_1) \right\} \nonumber \\
&=& \log(\rho_{3} \pi_{00}) + y(t) + \log \left\{1 + \exp(x(t))\right\} - \log\left\{\exp(-\lambda_1) + \exp(x(t)) \right\} \nonumber \\
&=& \log(\rho_{3} \pi_{00}) + y(t) + \log \left\{\frac{1 + \exp(x(t))}{\exp(-\lambda_1) + \exp(x(t))}\right\}, \nonumber \\
\log(\rho_{2} \pi_{01}) &=& z(t) - \log\left\{1 + \exp(y(t)) + \exp(z(t)) \right\} + x(t) + \lambda_2 \nonumber \\
&& - \: \log \left\{1 + \exp(x(t))\exp(\lambda_2) \right\} \nonumber \\
&=& \log(\rho_{3} \pi_{00}) + z(t) + \log \left\{\frac{1 + \exp(x(t))}{\exp(-\lambda_2) + \exp(x(t))}\right\}. \nonumber
\end{eqnarray}

Now, assume that there are two distinct points $\vec{\xi}$ and $\vec{\xi}^*$  ($\vec{\xi} \ne \vec{\xi}^*$) in the parameter space, such that the following equations (with obvious notation) hold:
\begin{eqnarray}
\rho_{1} \pi_{10} &=& \rho_{1}^* \pi_{10}^*, \label{iden:rho1} \\
\rho_{2} \pi_{01} &=& \rho_{2}^* \pi_{01}^*, \label{iden:rho2} \\
\rho_{3} \pi_{00} &=& \rho_{3}^* \pi_{00}^*, \label{iden:rho3}
\end{eqnarray}
for all $t \in \mathbb{R}$. By using (\ref{iden:rho3}), the equations (\ref{iden:rho1}) and (\ref{iden:rho2}) are equivalent to
\begin{eqnarray}
\lefteqn{y(t) - y^*(t)} \nonumber\\
&=& \log \left\{\frac{1 + \exp(x^*(t))}{\exp(-\lambda^*_1) + \exp(x^*(t))}\right\} - \log \left\{\frac{1 + \exp(x(t))}{\exp(-\lambda_1) + \exp(x(t))}\right\}, \label{exper:rho1} \\
\lefteqn{z(t) - z^*(t)} \nonumber \\
&=& \log \left\{\frac{1 + \exp(x^*(t))}{\exp(-\lambda^*_2) + \exp(x^*(t))}\right\} - \log \left\{\frac{1 + \exp(x(t))}{\exp(-\lambda_2) + \exp(x(t))}\right\}, \label{exper:rho2}
\end{eqnarray}
respectively. In (\ref{exper:rho1}) and (\ref{exper:rho2}) the left hand sides are straight lines. Thus, in order to (\ref{exper:rho1}) and (\ref{exper:rho2}) hold for all $t$,  the
right hand sides must be constants. If these constants were 0 (because $\lambda_1 = \lambda_1^*
= \lambda_2 = \lambda_2^* = 0$), then (\ref{iden:rho3}) would no longer hold for $\vec{\xi} \ne \vec{\xi}^*$ and all $t$. Alternatively,  the right hand sides of (\ref{exper:rho1}) and (\ref{exper:rho2}) are non-zero constants  if $\tau_{\pi_2} = \tau^*_{\pi_2} = 0$. Then,
as a consequence, (\ref{iden:rho3}) still is valid, for $\vec{\xi} \ne \vec{\xi}^*$ and all $t$, eventually if $\tau_{\rho_{21}} = \tau_{\rho_{21}}^* = 0$ and $\tau_{\rho_{22}} = \tau_{\rho_{22}}^* = 0$. This allows us to state that: if $\Pro(D_k | T) \ne \Pro(D_k)$, with $k  = 1,2$, then the considered model (with the particular choice for the functions $h$ and $f$) is identifiable, i.e., the joint probabilities $\Pro(D_1 = 1, D_2 = 0, V = 1| T = t)$, $\Pro(D_1 = 0, D_2 = 1, V = 1| T = t)$ and $\Pro(D_1 = 0, D_2 = 0, V = 1| T = t)$ are determined by a unique set of parameters. Of course, this claim can be easily extended to handle the presence of a covariate vector, $\vec{A}$.

\section{The proposal}
\subsection{VUS estimators}
\label{sec:vus_est}
Let $\rho_{k(v)}= \Pro(D_k = 1|V = v, T, \vec{A})$, for $k = 1, 2$ and $v = 0, 1$. It is easy to see, for instance, that
\begin{eqnarray}
\rho_{1(v)} &=& \frac{\Pro(V = v, D_1 = 1| D_2 = 0, T, \vec{A})}{\Pro(V = v| T, \vec{A})} \nonumber \\
&=& \frac{\Pro(V = v| D_1 = 1, D_2 = 0, T, \vec{A}) \Pro(D_1 = 1| T, \vec{A})}{\Pro(V = v| T, \vec{A})}. \nonumber
\end{eqnarray}
Hence, we can get, in particular,
\begin{eqnarray}
{\rho}_{1(0)} &=& \frac{(1 - {\pi}_{10}){\rho}_{1}}{(1 - {\pi}_{10}){\rho}_{1} + (1 - {\pi}_{01}){\rho}_{2} + (1 - {\pi}_{00}){\rho}_{3}}, \nonumber \\
{\rho}_{2(0)} &=& \frac{(1 - {\pi}_{01}){\rho}_{2}}{(1 - {\pi}_{10}){\rho}_{1} + (1 - {\pi}_{01}){\rho}_{2} + (1 - {\pi}_{00}){\rho}_{3}}, \nonumber \\
{\rho}_{3(0)} &=& \frac{(1 - {\pi}_{00}){\rho}_{3}}{(1 - {\pi}_{10}){\rho}_{1} + (1 - {\pi}_{01}){\rho}_{2} + (1 - {\pi}_{00}){\rho}_{3}}. \nonumber
\end{eqnarray}
Clearly, we also may consider quantities as
\begin{eqnarray}
\rho_{1(1)} &=&  \frac{{\pi}_{10}{\rho}_{1}}{{\pi}_{10}{\rho}_{1} + {\pi}_{01}{\rho}_{2} + {\pi}_{00}{\rho}_{3}}. \nonumber 
\end{eqnarray}
Then, we observe that
\begin{eqnarray}
\E(D_{1i}D_{2\ell}D_{3r}\I_{i\ell r}) &=& \E_{T, \vec{A}}\left\{\I_{i\ell r} \E (D_{1i}D_{2\ell}D_{3r} | T_i, \vec{A}_i, T_\ell, \vec{A}_\ell, T_r, \vec{A}_r) \right\}, \nonumber \\
&=& \E_{T, \vec{A}}\left\{\I_{i\ell r} \E (D_{1i}| T_i, \vec{A}_i)\E (D_{2\ell}|T_\ell, \vec{A}_\ell) \E (D_{3r}|T_r, \vec{A}_r) \right\}, \nonumber\\
&=& \E_{T, \vec{A}} \left(\rho_{1i}\rho_{2\ell}\rho_{3r} \I_{i\ell r} \right). \nonumber
\end{eqnarray}
Similarly, we have
\[
\E(D_{1i}D_{2\ell}D_{3r}) = \E_{T, \vec{A}} \left(\rho_{1i}\rho_{2\ell}\rho_{3r}\right),
\]
so that (\ref{org:vus}) can be rewritten as
\begin{equation}
\mu = \frac{\E_{T,\vec{A}} \left(\rho_{1i}\rho_{2\ell}\rho_{3r} \I_{i\ell r} \right)}{\E_{T, \vec{A}} \left(\rho_{1i}\rho_{2\ell}\rho_{3r}\right)}. 
\label{org:vus2}
\end{equation}

Equation (\ref{org:vus2}) suggests how to build estimators of VUS when some disease labels are missing in the sample: we can use suitable estimates $\hat{\rho}_{ki}$ to replace the $D_{ki}$'s in (\ref{nonp:vus}). Therefore, a FI estimator of VUS is simply
\begin{equation}
\hat{\mu}_{\mathrm{FI}} = \frac{\sum\limits_{i=1}^{n}\sum\limits_{\ell = 1, \ell \ne i}^{n} \sum\limits_{\stackrel{r = 1}{r \ne \ell, r\ne i}}^{n}\I_{i\ell r} \hat{\rho}_{1i}\hat{\rho}_{2\ell}\hat{\rho}_{3r}}{\sum\limits_{i = 1}^{n}\sum\limits_{\ell = 1, \ell \ne i}^{n} \sum\limits_{\stackrel{r = 1}{r \ne \ell, r \ne i}}^{n} \hat{\rho}_{1i}\hat{\rho}_{2\ell}\hat{\rho}_{3r}},
\label{fi:vus}
\end{equation} 
where $\hat{\rho}_{ki}$ ($k=1,2,3$ and $i=1,\ldots,n)$ are the estimated disease probabilities obtained from the disease model (\ref{dise:model}). 

Since $\E[V_i \rho_{k(1)i} + (1 - V_i)\rho_{k(0)i}|T, \vec{A}] = \rho_{ki}$, an alternative FI estimator of VUS could be obtained by replacing $D_{ki}$'s in (\ref{nonp:vus}) with the estimates $\tilde{D}_{ki, \mathrm{FI}} = V_i \hat{\rho}_{k(1)i} + (1 - V_i)\hat{\rho}_{k(0)i}$. Unlike FI approach, MSI estimator only replace the disease status $D_{ki}$ by the estimate $\hat{\rho}_{k(0)i}$ for unverified subjects. Define $D_{ki,\mathrm{MSI}} = V_i D_{ki} + (1 - V_i)\rho_{k(0)i}$ and let $\tilde{D}_{ki, \mathrm{MSI}}$ be the estimated version with $\rho_{k(0)i}$ replaced by $\hat{\rho}_{k(0)i}$, and
\begin{eqnarray}
\hat{\rho}_{1(0)i} &=& \frac{(1 - \hat{\pi}_{10i})\hat{\rho}_{1i}}{(1 - \hat{\pi}_{10i})\hat{\rho}_{ki} + (1 - \hat{\pi}_{01i})\hat{\rho}_{2i} + (1 - \hat{\pi}_{00i})\hat{\rho}_{3i}}, \nonumber \\
\hat{\rho}_{2(0)i} &=& \frac{(1 - \hat{\pi}_{01i})\hat{\rho}_{2i}}{(1 - \hat{\pi}_{10i})\hat{\rho}_{1i} + (1 - \hat{\pi}_{01i})\hat{\rho}_{2i} + (1 - \hat{\pi}_{00i})\hat{\rho}_{3i}}, \nonumber \\
\hat{\rho}_{3(0)i} &=& \frac{(1 - \hat{\pi}_{00i})\hat{\rho}_{3i}}{(1 - \hat{\pi}_{10i})\hat{\rho}_{1i} + (1 - \hat{\pi}_{01i})\hat{\rho}_{2i} + (1 - \hat{\pi}_{00i})\hat{\rho}_{3i}}. \nonumber
\end{eqnarray}
Here, $\hat{\pi}_{10i} = \widehat{\Pro}(V_i = 1|D_{1i} = 1, D_{2i} = 0, T_i, A_i)$, $\hat{\pi}_{01i} = \widehat{\Pro}(V_i = 1|D_{1i} = 0, D_{2i} = 1, T_i, A_i)$ and $\hat{\pi}_{00i} = \widehat{\Pro}(V_i = 1|D_{1i} = 0, D_{2i} = 0, T_i, A_i)$. Such estimates are 
derived from the verification model (\ref{veri:model:2}). Then, the MSI estimator of VUS 
is
\begin{equation}
\hat{\mu}_{\mathrm{MSI}} = \frac{\sum\limits_{i = 1}^{n}\sum\limits_{\ell = 1, \ell \ne i}^{n} \sum\limits_{\stackrel{r = 1}{r \ne \ell, r \ne i}}^{n}\I_{i\ell r}\tilde{D}_{1i,\mathrm{MSI}}\tilde{D}_{2\ell,\mathrm{MSI}}\tilde{D}_{3r,\mathrm{MSI}}}{\sum\limits_{i=1}^{n}\sum\limits_{\ell = 1, \ell \ne i}^{n} \sum\limits_{\stackrel{r = 1}{r \ne \ell, r \ne i}}^{n} \tilde{D}_{1i,\mathrm{MSI}}\tilde{D}_{2\ell,\mathrm{MSI}}\tilde{D}_{3r,\mathrm{MSI}}}.
\label{msi:vus}
\end{equation} 

In the IPW approach, instead, each observation in the subset of verified units is weighted by the inverse of the probability that the unit was selected for verification. Thus, the IPW estimator of VUS is 
\begin{equation}
\hat{\mu}_{\mathrm{IPW}} = \frac{\sum\limits_{i = 1}^{n}\sum\limits_{\ell = 1, \ell \ne i}^{n} \sum\limits_{\stackrel{r = 1}{r \ne \ell, r \ne i}}^{n}\I_{i\ell r} V_i V_\ell V_r D_{1i} D_{2\ell} D_{3r} \hat{\pi}_i^{-1} \hat{\pi}_\ell^{-1} \hat{\pi}_r^{-1}}{\sum\limits_{i = 1}^{n}\sum\limits_{\ell = 1, \ell \ne i}^{n} \sum\limits_{\stackrel{r = 1}{r \ne \ell, r \ne i}}^{n} V_i V_\ell V_r D_{1i} D_{2\ell} D_{3r} \hat{\pi}_i^{-1} \hat{\pi}_\ell^{-1}  \hat{\pi}_r^{-1}}.
\label{ipw:vus}
\end{equation}
Clearly, the estimates $\hat{\pi}_i$ also arise from the selection model (\ref{veri:model:2}). 

The last estimator is the pseudo doubly robust (PDR) estimator. We define
\[
D_{ki,\mathrm{PDR}} = \frac{V_i D_{ki}}{\pi_i} - \frac{\rho_{k(0)i}(V_i - \pi_i)}{\pi_i}.
\]
An estimated version, $\tilde{D}_{ki,\mathrm{PDR}}$, is obtained by entering the estimates $\hat{\pi}_i$ and $\hat{\rho}_{k(0)i}$ in the expression above. Then, the PDR estimator of VUS
is
\begin{equation}
\hat{\mu}_{\mathrm{PDR}} = \frac{\sum\limits_{i = 1}^{n}\sum\limits_{\ell = 1, \ell \ne i}^{n} \sum\limits_{\stackrel{r = 1}{r \ne \ell, r \ne i}}^{n}\I_{i\ell r}\tilde{D}_{1i,\mathrm{PDR}} \tilde{D}_{2\ell,\mathrm{PDR}} \tilde{D}_{3r,\mathrm{PDR}}} {\sum\limits_{i=1}^{n}\sum\limits_{\ell = 1, \ell \ne i}^{n} \sum\limits_{\stackrel{r = 1}{r \ne \ell, r \ne i}}^{n} \tilde{D}_{1i,\mathrm{PDR}} \tilde{D}_{2\ell,\mathrm{PDR}} \tilde{D}_{3r,\mathrm{PDR}}}.
\label{spe:vus}
\end{equation} 
The PDR estimator has the same nature as the SPE estimator discussed in \citet{toduc} under MAR assumption. However, under NI missing data mechanism it no longer has the doubly robust property. In fact, correct specification of both the verification model and the disease model is required for the PDR estimator to be consistent.

Note that all VUS estimators basically require maximum likelihood estimates of the parameters $\vec{\lambda}$, $\vec{\tau}_\pi$ and $\vec{\tau}_\rho$ of the working models (\ref{veri:model:2}) and (\ref{dise:model}).

\subsection{Asymptotic behavior}
\label{sec:asy_beha}
Let $\vec{\xi} = (\vec{\lambda}^\top, \vec{\tau}^\top_\pi, \vec{\tau}^\top_\rho)^\top$ be the nuisance parameter. Observe that the proposed VUS estimators can be found as solutions of appropriate estimating equations (solved along with the score equations). The estimating functions for FI, MSI, IPW and PDR estimators  have generic term (corresponding to a generic triplet of sample units), respectively, 
\begin{eqnarray}
G_{i\ell r, \mathrm{FI}}(\mu, \vec{\xi}) &=& \rho_{1i}(\vec{\tau}_\rho)\rho_{2\ell}(\vec{\tau}_\rho)\rho_{3r}(\vec{\tau}_\rho) \left(I_{i\ell r} - \mu \right), \nonumber \\
G_{i\ell r, \mathrm{MSI}}(\mu, \vec{\xi}) &=& D_{1i,\mathrm{MSI}}(\vec{\xi}) D_{2\ell, \mathrm{MSI}}(\vec{\xi}) D_{3r, \mathrm{MSI}}(\vec{\xi}) \left(I_{i\ell r} - \mu\right), \nonumber \\
G_{i\ell r, \mathrm{IPW}}(\mu, \vec{\xi}) &=& \frac{V_iV_\ell V_r D_{1i}D_{2\ell}D_{3r}}{\pi_i(\vec{\xi})\pi_\ell(\vec{\xi})\pi_k(\vec{\xi})} \left(I_{i\ell r} - \mu\right), \nonumber \\
G_{i\ell r,\mathrm{PDR}}(\mu, \vec{\xi}) &=& D_{1i, \mathrm{PDR}}(\vec{\xi}) D_{2\ell, \mathrm{PDR}}(\vec{\xi}) D_{3r, \mathrm{PDR}}(\vec{\xi}) \left(I_{i\ell r} - \mu\right). \nonumber
\end{eqnarray}
In the following, we will use the general notation $G_{i\ell r,*}(\mu, \vec{\xi})$, where the star stands for FI, MSI, IPW and PDR. 

Recall that the nuisance parameter $\vec{\xi}$ is estimated by maximizing the log--likelihood function (\ref{lg-like}). 
Let $\mathcal{S}_i(\vec{\xi})$ be the $i$--th subject's contribution to the score function,  $\mathcal{I}(\vec{\xi}) = -\E \left( \frac{\partial}{\partial \vec{\xi}^\top}\mathcal{S}_i(\vec{\xi}) \right)$ the Fisher information matrix for $\vec{\xi}$ and  $\hat{\vec{\xi}}$ the maximum likelihood estimator. Let $\mu_0$ be the true VUS value, and $\vec{\xi}_0 = (\vec{\lambda}^\top_0, \vec{\tau}^\top_{0\pi}, \vec{\tau}^\top_{0\rho})^\top$ the true value of $\vec{\xi}$. To give general theoretical results, we  assume that:
\begin{enumerate} [(C1)]
	\item the U--process
	\[
	U_{n,*}(\mu, \vec{\xi}) = \sqrt{n}\left\{G_{*}(\mu,\vec{\xi}) - e_*(\mu,\vec{\xi})\right\} 
	\]
	is stochastically equicontinuous, where
	\begin{eqnarray}
	G_{*}(\mu,\vec{\xi}) &=& \frac{1}{6n(n-1)(n-2)} \sum\limits_{i = 1}^{n}\sum\limits_{\ell = 1, \ell \ne i}^{n}\sum\limits_{\stackrel{r = 1}{r \ne \ell, r \ne i}}^{n} \bigg\{
	G_{i\ell r,*}(\mu,\vec{\xi}) + G_{ir \ell,*}(\mu,\vec{\xi}) \nonumber\\
	&& + \: G_{\ell ir,*}(\mu,\vec{\xi}) + G_{\ell r i,*}(\mu,\vec{\xi}) + G_{r i\ell ,*}(\mu,\vec{\xi}) + G_{r \ell i,*}(\mu,\vec{\xi}) \bigg\} \nonumber 
	\end{eqnarray}
	and
	\begin{eqnarray}
	e_*(\mu,\vec{\xi}) &=& \frac{1}{6} \E \bigg\{
	G_{i\ell  r,*}(\mu,\vec{\xi}) + G_{ir \ell ,*}(\mu,\vec{\xi}) + G_{\ell ir,*}(\mu,\vec{\xi}) + G_{\ell r i,*}(\mu,\vec{\xi}) \nonumber\\
	&& + \: G_{r i\ell ,*}(\mu,\vec{\xi}) + G_{r \ell i,*}(\mu,\vec{\xi}) \bigg\} \nonumber;
	\end{eqnarray}
	\item $e_*(\mu,\vec{\xi})$ is differentiable in $(\mu,\vec{\xi})$, and
	$\dfrac{\partial e_*(\mu,\vec{\xi}_0)}{\partial \mu}\Bigg|_{\mu = \mu_0}\ne 0$;
	\item $G_{*}(\mu,\vec{\xi})$ and $\dfrac{\partial G_{*}(\mu,\vec{\xi})}{\partial \vec{\xi}^\top}$ converges uniformly (in probability) to $e_*(\mu,\vec{\xi})$ and $\dfrac{\partial e_*(\mu,\vec{\xi})}{\partial \vec{\xi}^\top}$, respectively.
\end{enumerate}
We now state the two main results about consistency and asymptotic normality of the proposed estimators, whose proves are given in Appendix 1.
\begin{theorem}[Consistency]
	\label{thrm:1}
	Suppose that conditions (C1)--(C3) hold. Then, under the verification model (\ref{veri:model:2}) and the disease model (\ref{dise:model}),  $\hat{\mu}_* \stackrel{p}{\to} \mu_0$.
\end{theorem}
Recall that here the star indicates FI, MSI, IPW, and PDR.
\begin{theorem}[Asymptotic normality] \label{thrm:2}
	Under conditions (C1)--(C3), if the verification model (\ref{veri:model:2}) and the disease model (\ref{dise:model}) hold, then
	\begin{equation}
	\sqrt{n}\left(\hat{\mu}_{*} - \mu_0 \right) \stackrel{d}{\to} \mathcal{N}(0,\Lambda_*), \nonumber
	\end{equation}
	where $\Lambda_*$ is given in (\ref{lambdastar}).
\end{theorem}

It is worth noting that conditions (C1)--(C3) hold in our working model, which is based on (\ref{veri:model:2}), with $h(T, \vec{A}; \vec{\tau}_\pi) = \tau_{\pi_1} + \tau_{\pi_2}T + \vec{A}^\top \vec{\tau}_{\pi_3}$, and (\ref{dise:model}), with $f(T, \vec{A}; \vec{\tau}_{\rho_k}) = \tau_{\rho_{1k}} + \tau_{\rho_{2k}}T + \vec{A}^\top \vec{\tau}_{\rho_{3k}}$. In Appendix~1 we discuss how to obtain a consistent estimator of $\Lambda_*$. 

\section{Simulation study}
\label{sec:simula}
In this section, we provide empirical evidence, through simulation experiments, on the behavior of the proposed VUS estimators in finite samples. The number of replications in each simulation experiment is set to be 1000.

In the study, we consider two scenarios which correspond to quite different values of the true VUS. For both scenarios, we fix three sample sizes: 250, 500 and 1500.

In the first scenario, for each unit, we generate  the test result $T_i$ and a covariate $A_i$ from a bivariate normal distribution,
\[
(T_i,A_i) \sim \mathcal{N}_2\left(
\begin{pmatrix}
3.7 \\ 1.85
\end{pmatrix}, \begin{pmatrix}
3.71 & 1.36 \\ 1.36 & 3.13
\end{pmatrix}
\right).
\] 
The disease status $\vec{\mathcal{D}}_i$ is generated according to model (\ref{dise:model}) with $f(T, A;\vec{\tau}_{\rho_1}) = 4.6 - 3.3T - 6.4A$ and $f(T, A;\vec{\tau}_{\rho_2}) = 4 - 1.7T - 3.2A$. Then, the verification label $V_i$ is obtained according to model (\ref{veri:model:2}) with $h(T, A;\vec{\tau}_\pi) = 1 + 1.2T - 1.5A$ and $\lambda_1 = -2.5, \lambda_2 = -1$. Under such data generating process, $\theta_1 = 0.4$, $\theta_2 = 0.35$, $\theta_3 = 0.25$, and the verification rate is roughly $0.57$. The true VUS value is $0.791$. In the second scenario, we generate the test result and the covariate from independent normal distributions. Specifically, $T_i \sim \mathcal{N}(0.65, 1)$ and $A_i \sim \mathcal{N}(-0.3, 0.64)$. The disease status $\vec{\mathcal{D}}_i$ is generated according to model (\ref{dise:model}) with $f(T, A; \vec{\tau}_{\rho_1}) = 4.6 - 3.3T - 6.4A$ and $f(T, A; \vec{\tau}_{\rho_2}) = 4 - 1.7T - 3.2A$. Then, $V_i$ is obtained according to model (\ref{veri:model:2}) with $h(T,A; \vec{\tau}_\pi) = 1 + 1.2T - 1.5A$ and $\lambda_1 = -2.5, \lambda_2 = -1$. Under this setting, $\theta_1 = 0.55$, $\theta_2 = 0.32$, $\theta_3 = 0.13$, and the verification rate is roughly $0.58$. The true VUS value is $0.387$. 

Table \ref{tab:result1} contains  Monte Carlo means, Monte Carlo standard deviations and estimated standard deviations for the proposed VUS estimators (FI, MSI, IPW, PDR) in the two considered scenarios, at the chosen sample sizes. The table also reports the empirical coverages of the 95\% confidence intervals for the VUS, obtained through the
normal approximation approach applied to each estimator. To make a comparison, Table \ref{tab:result1} also gives the results for the semiparametric efficient estimator (SPE) discussed in \citet{toduc}, whose realizations are obtained, in all experiments, under the MAR
assumption, i.e., by setting $\lambda_1 = \lambda_2 = 0$ in model (\ref{veri:model:2}). The comparison allows us to evaluate the possible impact of an incorrect hypothesis MAR on the most robust estimator among those, FI, MSI, IPW and SPE, which are built to work under ignorable missing data mechanism (see \citet{toduc}).

\begin{table}[htbp]
	\caption{Monte Carlo means (MCmean),  relative bias (Bias), Monte Carlo
		standard deviations (MCds) and estimated standard deviations (Esd) for the proposed VUS estimators, and the SPE estimator under MAR assumption. CP denotes Monte Carlo coverages for the  95\% confidence intervals, obtained through the
		normal approximation approach applied to each estimator.}
	\begin{center}
		\begin{small}
			\begin{tabular}{c c l r r r c c}
				\toprule
				& \centering Sample size & Estimators & MCmean  &  Bias(\%)  &  MCsd  & Esd &  CP(\%)\\
				\midrule
				\multirow{17}{1.8 cm}{\centering Scenario I: VUS = 0.791 } & \multirow{5}{*}{$n = 250$} & FI & 0.772 & -2.4 & 0.056 & 0.050 & 89.9 \\ 
				& & MSI & 0.770 & -2.7 & 0.057 & 0.051 & 90.6 \\ 
				& & IPW & 0.770 & -2.6 & 0.070 & 0.061 & 88.1 \\ 
				& &  PDR & 0.766 & -3.2 & 0.085 & 0.075 & 90.8 \\ 
				& &  SPE (MAR) & 0.771 & -2.5 & 0.073 & 0.138 & 93.2 \\ 
				& & & & & & & \\
				\multirow{11}{1 cm}{\ } & \multirow{5}{*}{$n = 500$} & FI  & 0.783 & -1.0 & 0.035 & 0.032 &  93.3 \\
				& & MSI & 0.782 & -1.1 & 0.036 & 0.033 &  93.4 \\
				& & IPW & 0.782 & -1.2 & 0.047 & 0.042 &  92.2 \\
				& & PDR & 0.782 & -1.2 & 0.053 & 0.058 &  94.0 \\
				& & SPE (MAR) & 0.771 & -2.6 & 0.047 & 0.040 &  93.0 \\
				& & & & & & & \\
				& \multirow{5}{*}{$n = 1500$} & FI & 0.790 & -0.2 & 0.016 & 0.016 &  95.0 \\
				& & MSI & 0.789 & -0.2 & 0.016 & 0.016 & 95.2 \\
				& & IPW & 0.788 & -0.3 & 0.025 & 0.024 & 94.4 \\
				& & PDR & 0.789 & -0.3 & 0.025 & 0.024 & 95.2 \\
				& & SPE (MAR) & 0.771 & -2.5 & 0.027 & 0.025 & 89.4 \\
				\midrule
				\multirow{17}{1.8 cm}{\centering Scenario II: VUS = 0.387 } & \multirow{5}{*}{$n = 250$} &
				FI & 0.368 & -5.0 & 0.064 & 0.057 & 87.4 \\ 
				& & MSI & 0.367 & -5.2 & 0.065 & 0.059 & 87.9 \\ 
				& & IPW & 0.377 & -2.6 & 0.084 & 0.074 & 87.6 \\ 
				& & PDR & 0.369 & -4.6 & 0.086 & 0.075 & 89.5 \\ 
				& &  SPE (MAR) & 0.346 & -10.6 & 0.063 & 0.058 & 84.5 \\ 
				& & & & & & & \\
				\multirow{11}{1 cm}{\ } & \multirow{5}{*}{$n = 500$} & FI  & 0.379 & -2.0 & 0.045 & 0.041 & 90.9 \\
				& & MSI & 0.379 & -2.1 & 0.046 & 0.042 &  91.3 \\
				& & IPW & 0.380 & -1.8 & 0.060 & 0.056 &  91.2 \\
				& & PDR & 0.381 & -1.6 & 0.060 & 0.053 &  92.0 \\
				& & SPE (MAR) & 0.345 & -10.8 & 0.044 & 0.042 & 76.5 \\
				& & & & & & & \\
				& \multirow{5}{*}{$n = 1500$} & FI &  0.388 & 0.2 & 0.023 & 0.022 & 94.2 \\
				& & MSI & 0.388 &  0.2 & 0.023 & 0.023 &  94.3 \\
				& & IPW & 0.388 &  0.3 & 0.034 & 0.032 &  94.9 \\
				& & PDR & 0.389 &  0.4 & 0.033 & 0.029 &  93.2 \\
				& & SPE (MAR) & 0.346 & -10.7 & 0.026 & 0.025 &  76.5 \\
				\bottomrule
			\end{tabular}
		\end{small}
	\end{center}
	\label{tab:result1}
\end{table}

Overall, simulation results are consistent with our theoretical findings and show the usefulness of the proposed estimators, which also arises from the comparison with the SPE estimator used improperly. The results also show a good behavior of the estimated standard deviations, which are generally close to the corresponding Monte Carlo values. In general, FI and MSI estimators seem to be more efficient than IPW and PDR estimators. However, for all estimators, acceptable bias levels and sufficiently accurate associated confidence intervals seem to require a large sample size (at least 500, and, prudently, even higher). 

This issue of poor accuracy has already been noted by several authors, including \citet{liu}, in the context of two-class classification problems. In our experience, the trouble appears to arise because of a bad behavior of the maximum likelihood estimates  in the verification  and disease models. If the sample size is not large enough, the data do not contain enough information to effectively estimate the parameters $\vec{\lambda}$, $\vec{\tau}_\pi,$ $\vec{\tau}_{\rho_1}$ and $\vec{\tau}_{\rho_2}$. It seems particularly difficult to get good estimates of nonignorable parameters.

Table \ref{tab:mle}, giving  the Monte Carlo means for the maximum likelihood estimators of the elements of $\vec{\lambda}$, $\vec{\tau}_\pi,$ $\vec{\tau}_{\rho_1}$ and $\vec{\tau}_{\rho_2}$, for the three considered sample sizes, allows us to look at the bias of the estimators. More importantly, Figure 1 and Figure 2 (which refer to scenario I and II, respectively) graphically depict values of the estimates of $\lambda_1$, $\lambda_2$ and $\tau_{\pi_1}$ obtained in the thousand replications, for each sample size. The plots clearly show the great variability of the maximum likelihood estimates at lower sample sizes, with many values dramatically different from the corresponding target values. With larger sample size, this phenomenon almost completely vanishes, the maximum likelihood estimators behave pretty well, with a positive impact on the behavior of the VUS estimators.

\begin{table}[htbp]
	\caption{Monte Carlo means (MCmean) for the maximum likelihood estimators of the elements of nuisance parameters
		$\vec{\lambda}$, $\vec{\tau}_\pi$, $\vec{\tau}_{\rho_1}$ and $\vec{\tau}_{\rho_2}$.}
	\label{tab:mle}
	\begin{center}
		\begin{tabular}{c r r r r | r r r r}
			\toprule
			& \multicolumn{4}{c|}{Scenario I} & \multicolumn{4}{c}{Scenario II} \\
			\midrule
			& True & \multicolumn{3}{c|}{\, MCmean} & True &
			\multicolumn{3}{c}{\, MCmean} \\
			&  & $n=250$ & $n=500$ & $n=1500$ & & $n=250$ & $n=500$ & $n=1500$ \\
			\midrule
			$\lambda_1$ & -2.00 & -1.01 & -1.76 & -1.95 & -2.50 & -2.09 & -2.30 &
			-2.50 \\
			$\lambda_2$ & -1.00 & -0.45 & -0.87 & -0.98 & -1.00 & -0.99 & -0.96 &
			-0.97 \\
			$\tau_{\pi_1}$ & 2.00 & 1.25 & 1.80 & 1.95 & 1.00 & 1.17 & 1.00 & 1.00 \\
			$\tau_{\pi_2}$ & 0.50 & 0.65 & 0.55 & 0.51 & 1.20 & 1.39 & 1.28 & 1.22 \\
			$\tau_{\pi_3}$ & -1.20 & -1.24 & -1.22 & -1.21 & -1.50 & -1.25 & -1.40 &
			-1.51 \\
			$\tau_{\rho_{11}}$ & 15.00 & 15.53 & 15.28 & 15.10 & 4.60 & 4.44 & 4.58 &
			4.66 \\
			$\tau_{\rho_{21}}$ & -3.30 & -3.41 & -3.36 & -3.32 & -3.30 & -3.29 & -3.33
			& -3.34 \\
			$\tau_{\rho_{31}}$ & -0.70 & -0.89 & -0.78 & -0.72 & -6.40 & -6.94 & -6.70
			& -6.48 \\
			$\tau_{\rho_{12}}$ & 9.50 & 10.03 & 9.71 & 9.57 & 4.00 & 4.12 & 4.11 &
			4.05 \\
			$\tau_{\rho_{22}}$ & -1.70 & -1.79 & -1.73 & -1.71 & -1.70 & -1.77 & -1.76
			& -1.73 \\
			$\tau_{\rho_{32}}$ & -0.30 & -0.40 & -0.34 & -0.31 & -3.20 & -3.62 & -3.42
			& -3.25 \\
			\bottomrule
		\end{tabular}
	\end{center}
\end{table}

\begin{figure}[htbp]
	\begin{center}
		\includegraphics[width = 1\textwidth]{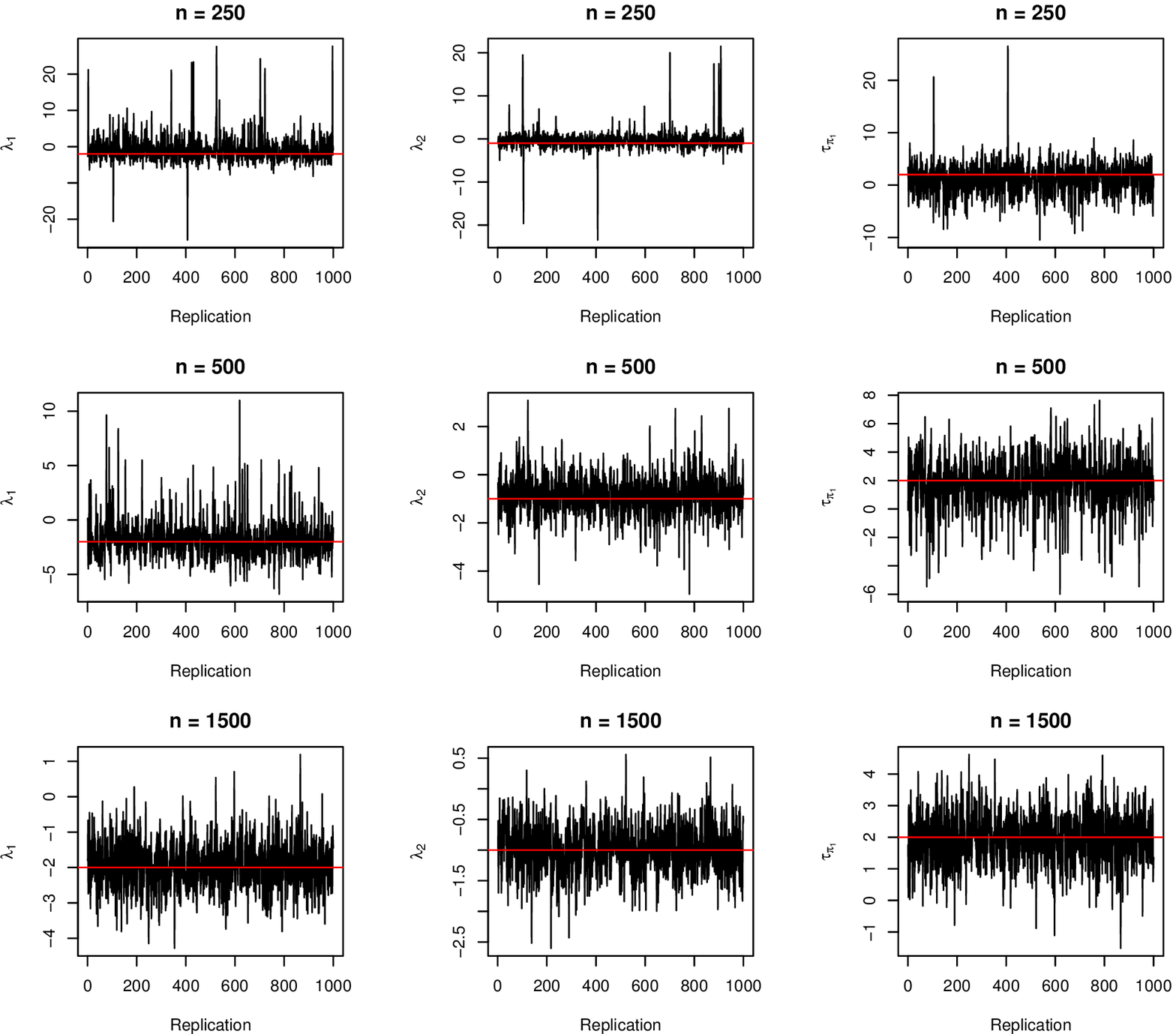}
		\caption{Values of the estimates 
			of $\lambda_1,$ $\lambda_2,$ and $\tau_{\pi_1}$, obtained in thousand Monte Carlo replications (Scenario I), for each sample size. The horizontal lines indicate the true parameter values.}
		\label{fig:1}
	\end{center}
\end{figure}

\begin{figure}[htbp]
	\begin{center}
		\includegraphics[width = 1\textwidth]{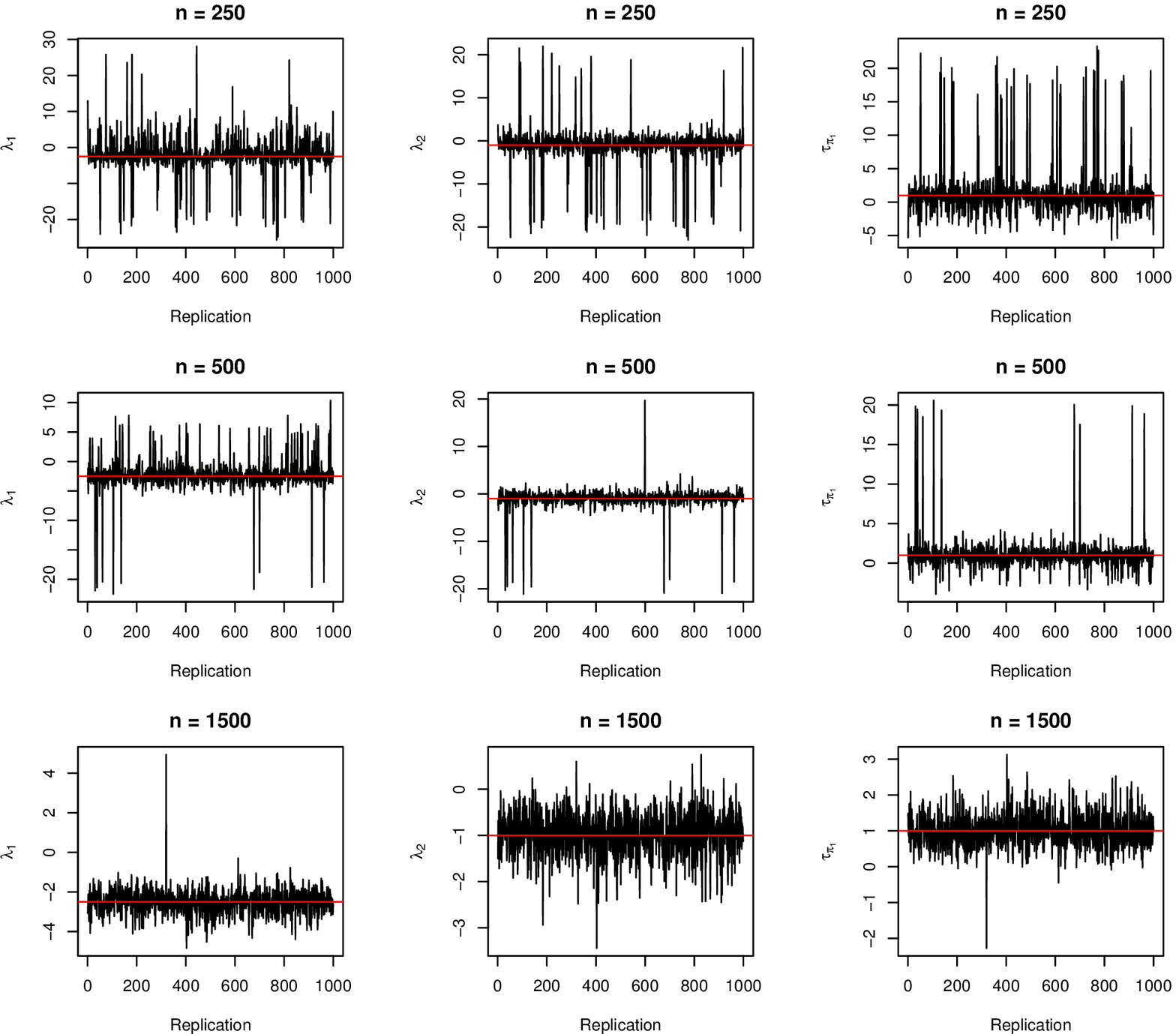}
		\caption{Values of the estimates of $\lambda_1,$ $\lambda_2,$ and $\tau_{\pi_1}$, obtained in thousand Monte Carlo replications (Scenario II), for each sample size. The horizontal lines indicate the true parameter values.}
		\label{fig:2}
	\end{center}
\end{figure}

\section{An illustration}
To illustrate the application of our proposed methods, we used data from Alzheimer's Disease Neuroimaging Initiative (ADNI, adni.loni.usc.edu). ADNI was launched in 2003 as a public-private partnership with the primary goal of finding suitable diagnostic tests or biomarkers for early detection and tracking of the Alzheimer's disease (see www.adni-info.org for up-to-date information).

Study subjects are classified in one of three classes, i.e., cognitively normal (CN), mild cognitive impairment (MCI), and Alzheimer's disease (AD) on the basis of neuropsychological tests. Various clinical, imaging, genetic and biochemical markers are also available. Among them, we consider cerebrospinal fluid (CSF) tau protein as the primary marker, and amyloid beta 1-42 (A$\beta$1-42) as a covariate. The full dataset in our illustration refers to 1209 subjects (CN: 363, MCI: 618, AD: 228 at baseline visit). Exploratory analysis suggests that high values of CFS tau protein are associated with severe disease status.

Here, we want to evaluate  the accuracy of CSF tau protein as a marker for the Alzheimer's disease assuming  nonignorable  missingness in the disease status of some patients.  To this aim, we induce  missingness by randomly selecting patients from the three classes in different proportions. Then, about 66\%  and 16\% patients in AD and MCI classes, respectively, are given a missing disease status. No missing status is set in the CN class. Overall, in the resulting dataset, for 76.4\% of the subjects the true disease status is present. For convenience, values of the considered variables, i.e., CSF tau protein and amyloid beta 1-42, are standardized. In what follows, the standardized variables will be denoted as $T$ and $A,$ respectively.

We focus on VUS estimation of marker $T$ by applying the FI, MSI, IPW and PDR estimators proposed in Section~3.  For the sake of comparison, bias--corrected VUS estimators under MAR assumption (\citet{toduc}), are also employed. Estimates under the MAR assumption are obtained through the R package \texttt{bcROCsurface} (\citet{toduc2017bcrocsurface}), in which the verification model is fitted by using logistic regression with $T$ and $A$ as covariates, and the disease model is fitted by means of multinomial logistic regression with $T$ and $A$ as regressors. Moreover, we also compute the VUS estimate based on the full dataset (Full estimate).

The maximum likelihood estimate of the non-ignorable parameter $\boldsymbol\lambda = (\lambda_1, \lambda_2)^\top$ in the verification model (3) is  $(18.766, 3.344)^\top$. The log-likelihood ratio test for the hypothesis $H_0: \boldsymbol\lambda = \boldsymbol{0}$ versus $H_{\mathrm{A}}: \boldsymbol\lambda \ne \boldsymbol{0}$ achieves a $p$-value of 0.0066, indicating that the non-ignorable effect is significant.

\begin{table}[htpb]
	\begin{center}
		\begin{threeparttable}[b]
		\caption{ADNI dataset. Estimates of VUS for the standardized CSF tau protein, associated estimated standard deviations (SD) and 95\% confidence intervals. SD for Full estimated is obtained via bootstrap resampling.}
		\label{tab:vus_adni}
		\begin{tabular}{l l c c c}
			\toprule
			& & VUS Estimate & SD & 95\% confidence interval \\
			\midrule
			Full & & 0.338 & 0.019 & (0.301, 0.375) \\
			\midrule
			\multirow{4}{1.5cm}{NI} & FI & 0.335 & 0.016 & (0.303, 0.368) \\
			& MSI & 0.333 & 0.018 & (0.298, 0.368) \\
			& IPW & 0.352 & 0.024 & (0.305, 0.398) \\
			& PDR & 0.344 & 0.021 & (0.302, 0.384) \\
			\midrule
			\multirow{4}{2cm}{MAR} & FI & 0.369 & 0.031 & (0.309, 0.429) \\
			& MSI & 0.366 & 0.031 & (0.304, 0.427) \\
			& IPW & 0.373 & 0.034 & (0.307, 0.440) \\
			& SPE & 0.362 & 0.033 & (0.297, 0.426) \\
			\bottomrule
		\end{tabular}
	\begin{scriptsize}
	\end{scriptsize}
\end{threeparttable}
	\end{center}
\end{table}

Table \ref{tab:vus_adni} shows  the Full and the bias-corrected NI and MAR estimates of VUS. The table also gives the estimated standard deviations, and approximated 95\% confidence intervals. For the Full estimate, the bootstrap standard deviation is reported (250 bootstrap replications). Standard deviations of NI and MAR estimates are obtained by using asymptotic theory. As expected,  taking the Full estimate as a benchmark, MAR estimators seem to overestimate the VUS, whereas the NI estimators appear to perform better, in particular FI and MSI. Consistently with simulation results in Table~1, the NI bias-corrected FI and MSI estimators appear also to be more efficient than the IPW and PDR estimators.  
Taking into account that the logistic model (3) used for the verification process does not reflect the induced verification mechanism, this seems to suggest that  the NI bias-corrected FI and MSI estimators are less sensitive to possible misspecifications of the verification process. 

\section{Conclusion}
\label{sec:conclu}
In this paper, we have proposed four bias--corrected estimators of VUS under NI missing data mechanism. The estimators are obtained by a likelihood--based approach, which uses the verification model (\ref{veri:model:2}) together with the disease model (\ref{dise:model}). The identifiability of the joint model is proved, and hence, the nuisance parameters can be estimated by maximizing the log--likelihood function or solving the score equations. Consistency and asymptotic normality of the proposed FI, MSI, IPW and PDR estimators are established, and variance estimation is  discussed. 

The proposed VUS estimators are pretty easy to implement and require the use of some numerical routine to maximize the log--likelihood function (or to solve the score equations). Our simulation results show their usefulness, whilst confirming the evidence emerging in the two--class case, according to which a reasonable large sample size is necessary  to make sufficiently accurate inference. In practice, among FI, MSI, IPW and PDR estimators, we would reccommend FI and MSI estimators thanks to their greater efficiency.

The poor accuracy problem seems to be related to an intrinsic difficulty of the maximum likelihood method in providing accurate estimates of the parameters of the disease and verification models, in particular of the nonignorable parameters. Overcoming this drawback is a stimulating challenge and deserves further investigation.

\begin{acknowledgements}
The authors thank the Alzheimers Disease Neuroimaging
Initiative research group for kindly permitting access to the data analyzed in this paper.
Data collection and sharing for this project was funded by the Alzheimer's Disease
Neuroimaging Initiative (ADNI) (National Institutes of Health Grant U01 AG024904)
and DOD ADNI (Department of Defense award number W81XWH-12-2-0012). ADNI is
funded by the National Institute on Aging, the National Institute of Biomedical Imaging and
Bioengineering, and through generous contributions from the following: AbbVie, Alzheimers
Association; Alzheimers Drug Discovery Foundation; Araclon Biotech; BioClinica, Inc.;
Biogen; Bristol-Myers Squibb Company; CereSpir, Inc.; Eisai Inc.; Elan Pharmaceuticals, Inc.;
Eli Lilly and Company; EuroImmun; F. Hoffmann-La Roche Ltd and its affiliated company
Genentech, Inc.; Fujirebio; GE Healthcare; IXICO Ltd.; Janssen Alzheimer Immunotherapy
Research \& Development, LLC.; Johnson \& Johnson Pharmaceutical Research \& Development
LLC.; Lumosity; Lundbeck; Merck \& Co., Inc.; Meso Scale Diagnostics, LLC.; NeuroRx
Research; Neurotrack Technologies; Novartis Pharmaceuticals Corporation; Pfizer Inc.; Piramal
Imaging; Servier; Takeda Pharmaceutical Company; and Transition Therapeutics. The Canadian
Institutes of Health Research is providing funds to support ADNI clinical sites in Canada.
Private sector contributions are facilitated by the Foundation for the National Institutes of Health
(www.fnih.org). The grantee organization is the Northern California Institute for Research and
Education, and the study is coordinated by the Alzheimer?s Disease Cooperative Study at the
University of California, San Diego. ADNI data are disseminated by the Laboratory for Neuro
Imaging at the University of Southern California.
\end{acknowledgements}

\bibliographystyle{spbasic}      
\bibliography{ToDuc-Chiogna-Adimari}   

\appendix
\section*{Appendix 1}
{\bf Proves}\\~\\
{\it Proof of Theorem 1.}
	We can show that $\E\{G_{i\ell r,*}(\mu_0,\vec{\xi}_0)\} = 0$ (see the Appendix 2). Then $e_*(\mu_0,\vec{\xi}_0) = 0$, and, by condition (C2) and an application of implicit function theorem, there exists a neighborhood of $\vec{\xi}_0$ in which a continuously differentiable function, $m(\vec{\xi})$, is uniquely defined such that $m(\vec{\xi}_0) = \mu_0$ and $e_*(m(\vec{\xi}),\vec{\xi}) = 0$. Since the maximum likelihood estimator $\hat{\vec{\xi}}$ is consistent, 
	i.e., $\hat{\vec{\xi}} \stackrel{p}{\to} \vec{\xi}_0$, we have that $\tilde\mu_* = m(\hat{\vec{\xi}})\stackrel{p}{\to} \mu_0$.
	On the other hand, $G_*(\hat\mu_*, \hat{\vec{\xi}}) = 0$ and condition (C3) implies that $e_*(\hat\mu_*,\hat{\vec{\xi}})\stackrel{p}{\to} 0$.
	Thus, $\hat{\mu}_* \stackrel{p}{\to} \tilde\mu_*$.
	\qed

\vspace{0.3cm}
\noindent
{\it Proof of Theorem 2.}
	We have
	\begin{eqnarray}
	0 &=& \sqrt{n}G_{*}(\hat{\mu}_{*},\hat{\vec{\xi}}) \nonumber\\
	0 &=& \sqrt{n}G_{*}(\hat{\mu}_{*},\hat{\vec{\xi}}) + \sqrt{n}e_*(\hat{\mu}_{*},\hat{\vec{\xi}}) - \sqrt{n}e_*(\hat{\mu}_{*},\hat{\vec{\xi}}).\nonumber
	\end{eqnarray}
	Since $e_*(\mu_0,\vec{\xi}_0) = 0$, we get
	\begin{eqnarray}
	0 &=& \sqrt{n}G_{*}(\hat{\mu}_{*},\hat{\vec{\xi}}) + \sqrt{n}e_*(\hat{\mu}_{*},\hat{\vec{\xi}}) - \sqrt{n}e_*(\hat{\mu}_{*},\hat{\vec{\xi}}) + \sqrt{n}e_*(\mu_0,\vec{\xi}_0) - \sqrt{n}e_*(\mu_0,\vec{\xi}_0)\nonumber \\
	\ &=& \sqrt{n}\left\{G_{*}(\hat{\mu}_{*},\hat{\vec{\xi}}) - e_*(\hat{\mu}_{*},\hat{\vec{\xi}}) \right\} + \sqrt{n}\left\{ e_*(\hat{\mu}_{*},\hat{\vec{\xi}}) - e_*(\mu_0,\vec{\xi}_0)\right\} + \sqrt{n}e_*(\mu_0,\vec{\xi}_0)\nonumber \\
	&& - \: \sqrt{n}G_{*}(\mu_0,\vec{\xi}_0) + \sqrt{n}G_{*}(\mu_0,\vec{\xi}_0) \nonumber\\
	\ &=& \left[\sqrt{n}\left\{G_{*}(\hat{\mu}_{*},\hat{\vec{\xi}}) - e_*(\hat{\mu}_{*},\hat{\vec{\xi}}) \right\} - \sqrt{n}\left\{G_{*}(\mu_0,\vec{\xi}_0) - e_*(\mu_0,\vec{\xi}_0)\right\}\right]\nonumber \\
	&& + \:  \sqrt{n}\left\{ e_*(\hat{\mu}_{*},\hat{\vec{\xi}}) - e_*(\mu_0,\vec{\xi}_0)\right\} + \sqrt{n}G_{*}(\mu_0,\vec{\xi}_0) \nonumber.
	\end{eqnarray}
	Condition (C1) implies that the first term in right hand side of the last identity is $o_p(1)$. Using the Taylor expansion, we have
	\begin{eqnarray}
	0 &=& o_p(1) + \sqrt{n}\left\{ e_*(\hat{\mu}_{*},\hat{\vec{\xi}}) - e_*(\mu_0,\vec{\xi}_0)\right\} + \sqrt{n}G_{*}(\mu_0,\vec{\xi}_0) \nonumber \\
	\ &=& o_p(1) + \sqrt{n}(\hat{\mu}_{*} - \mu_0) \frac{\partial e_*(\mu,\vec{\xi}_0)}{\partial \mu}\Bigg|_{\mu = \mu_0} \nonumber\\
	&& + \: \sqrt{n}(\hat{\vec{\xi}} - \vec{\xi}_0)\frac{\partial e_*(\mu_0,\vec{\xi})}{\partial \vec{\xi}^\top}\Bigg|_{\vec{\xi} = \vec{\xi}_0} + \sqrt{n}G_{*}(\mu_0,\vec{\xi}_0) 
	\label{taylor}.
	\end{eqnarray}
	It is straightforward to show that
	\[
	\frac{\partial e_*(\mu,\vec{\xi}_0)}{\partial \mu}\Bigg|_{\mu = \mu_0} = - \Pro(D_1 = 1) \Pro(D_2 = 1) \Pro(D_3 = 1) = - \theta_1 \theta_2 \theta_3.
	\]
	By standard results on the limit distribution of U-statistics \citep[Theorem 12.3, Chap. 12]{van},
	\[
	\sqrt{n}U_{n,*}(\mu_0,\vec{\xi}_0)=
	\sqrt{n}\left\{G_{*}(\mu_0,\vec{\xi}_0) - e_*(\mu_0,\vec{\xi}_0)\right\} = \sqrt{n}G_{*}(\mu_0,\vec{\xi}_0) \stackrel{p}{\to} \sqrt{n}\tilde{G}_{*}(\mu_0,\vec{\xi}_0),
	\] 
	where $\sqrt{n}\tilde{G}_{*}(\mu,\vec{\xi})$ is the projection of $U_{n,*}$ onto the set of all statistics of the form 
	\begin{eqnarray}
	\sqrt{n}\tilde{G}_{n,*}(\mu,\vec{\xi}) &=& \frac{1}{2\sqrt{n}}\sum_{i=1}^{n} \E \bigg\{ G_{i\ell  r,*}(\mu,\vec{\xi}) + G_{ir \ell ,*}(\mu,\vec{\xi}) + G_{\ell ir,*}(\mu,\vec{\xi}) \nonumber\\
	&& + \: G_{\ell r i,*}(\mu,\vec{\xi}) + G_{r i\ell ,*}(\mu,\vec{\xi}) + G_{r \ell i,*}(\mu,\vec{\xi}) \big|O_i \bigg\} \nonumber
	\end{eqnarray}
	for $\ell \ne i$ and $r \ne \ell, r \ne i$. For the maximum likelihood estimator $\hat{\vec{\xi}}$, we can write
	\begin{eqnarray}
	\sqrt{n}\left(\hat{\vec{\xi}} - \vec{\xi}_0\right) &=& \frac{1}{\sqrt{n}}\left[-\frac{\partial \E \left\{\mathcal{S}_i(\vec{\xi})\right\}}{\partial \vec{\xi}^\top}\Bigg|_{\vec{\xi} = \vec{\xi}_0}\right]^{-1}\sum_{i=1}^{n}\mathcal{S}_i(\vec{\xi}_0) + o_p(1) \nonumber \\
	&=& \frac{1}{\sqrt{n}}\mathcal{I}(\vec{\xi})^{-1} \sum_{i=1}^{n}\mathcal{S}_i(\vec{\xi}_0) + o_p(1). \nonumber
	\end{eqnarray}
	Hence, from (\ref{taylor}),
	\begin{eqnarray}
	\lefteqn{\theta_1 \theta_2 \theta_3 \sqrt{n}(\hat{\mu}_{*} - \mu_0)} \nonumber \\
	&=& o_p(1) + \frac{1}{\sqrt{n}} \frac{\partial e_*(\mu_0,\vec{\xi})}{\partial \vec{\xi}^\top}\Bigg|_{\vec{\xi} = \vec{\xi}_0} \mathcal{I}(\vec{\xi})^{-1} \sum_{i=1}^{n}\mathcal{S}_i(\vec{\xi}_0) \nonumber \\
	&& + \: \frac{1}{2\sqrt{n}}\sum_{i=1}^{n} \E \bigg\{ G_{i\ell r,*}(\mu_0,\vec{\xi}_0) + G_{ir \ell ,*}(\mu_0,\vec{\xi}_0) + G_{\ell ir,*}(\mu_0,\vec{\xi}_0) \nonumber\\
	&& + \: G_{\ell r i,*}(\mu_0,\vec{\xi}_0) + G_{r i\ell ,*}(\mu_0,\vec{\xi}_0) + G_{r \ell i,*}(\mu_0,\vec{\xi}_0) \big|O_i \bigg\}\nonumber \\
	&=& o_p(1) + \frac{1}{\sqrt{n}}\sum_{i=1}^{n} \Bigg[ \frac{\partial e_*(\mu_0,\vec{\xi})}{\partial \vec{\xi}^\top}\Bigg|_{\vec{\xi} = \vec{\xi}_0} \mathcal{I}(\vec{\xi})^{-1} \mathcal{S}_i(\vec{\xi}_0) \nonumber \\
	&& + \: \frac{1}{2} \E \bigg\{ G_{i\ell r,*}(\mu_0,\vec{\xi}_0) + G_{ir \ell ,*}(\mu_0,\vec{\xi}_0) + G_{\ell ir,*}(\mu_0,\vec{\xi}_0) \nonumber\\
	&& + \: G_{\ell r i,*}(\mu_0,\vec{\xi}_0) + G_{r i\ell ,*}(\mu_0,\vec{\xi}_0) + G_{r \ell i,*}(\mu_0,\vec{\xi}_0) \big|O_i \bigg\} \Bigg] \label{qi} \\
	&=& o_p(1) + \frac{1}{\sqrt{n}}\sum_{i=1}^{n}Q_{i,*}(\mu_0,\vec{\xi}_0) = o_p(1) + \frac{1}{\sqrt{n}} Q_*(\mu_0,\vec{\xi}_0). \nonumber
	\end{eqnarray}
	Note that the observed data $O_i$ are i.i.d, then $Q_{i,*}(\mu_0,\vec{\xi}_0)$ are also i.i.d. In addition, we easily show that
	\begin{eqnarray}
	0 &=& \E \Bigg[\E \bigg\{ G_{i\ell r,*}(\mu_0,\vec{\xi}_0) + G_{ir \ell ,*}(\mu_0, \vec{\xi}_0) + G_{\ell ir,*}(\mu_0,\vec{\xi}_0) + G_{\ell r i,*}(\mu_0, \vec{\xi}_0) \nonumber\\
	&& + \: G_{r i\ell ,*}(\mu_0, \vec{\xi}_0) + G_{r \ell i,*}(\mu_0, \vec{\xi}_0) \big|O_i \bigg\} \Bigg] \nonumber.
	\end{eqnarray}
	Therefore, $\E \{Q_{i,*} (\mu_0,\vec{\xi}_0)\} = 0$, and  $\frac{1}{\sqrt{n}} Q_* (\mu_0,\vec{\xi}_0) \stackrel{d}{\to} \mathcal{N}(0, \V \left\{Q_{i,*} (\mu_0,\vec{\xi}_0)\right\})$ by the Central Limit Theorem. It follows that 
	\[
	\sqrt{n}\left(\hat{\mu}_{*} - \mu_0 \right) \stackrel{d}{\to} \mathcal{N}\left(0, \Lambda_*\right),
	\]
	where
\begin{eqnarray}
	\Lambda_* = \frac{\V \left\{Q_{i,*} (\mu_0,\vec{\xi}_0)\right\}}{\theta_1^2\theta_2^2\theta_3^2}.
\label{lambdastar}
\end{eqnarray}
	\qed

\noindent
{\bf Variance estimation}\\~\\
%
Under condition (C3), a consistent estimator of $\Lambda_*$ can be obtained as
\begin{equation}
\hat{\Lambda}_* = \frac{\V \left\{\hat{Q}_{i,*} (\hat{\mu}_{*},\hat{\vec{\xi}})\right\}}{\hat{\theta}_{1,*}^2 \hat{\theta}_{2,*}^2 \hat{\theta}_{3,*}^2} = \frac{\frac{1}{n - 1} \sum\limits_{i=1}^{n}\hat{Q}_{i,*}^2(\hat{\mu}_{*},\hat{\vec{\xi}})}{\hat{\theta}_{1,*}^2 \hat{\theta}_{2,*}^2 \hat{\theta}_{3,*}^2},
\label{asy:var:vus}
\end{equation}
where $\hat{\theta}_{k,*}$ are the estimates of the disease probabilities, $\theta_{k}$ for $k = 1,2,3$. Specifically, $\hat{\theta}_{k,\mathrm{FI}} = \frac{1}{n}\sum\limits_{i=1}^{n} \hat{\rho}_{ki}$, $\hat{\theta}_{k,\mathrm{MSI}} = \frac{1}{n}\sum\limits_{i=1}^{n} \tilde{D}_{ki,\mathrm{MSI}}$, $\hat{\theta}_{k,\mathrm{IPW}} = \sum\limits_{i=1}^{n} V_iD_{ki}\hat{\pi}_i^{-1} / \sum\limits_{i=1}^{n} V_i\hat{\pi}_i^{-1}$ and $\hat{\theta}_{k,\mathrm{PDR}} = \frac{1}{n}\sum\limits_{i=1}^{n} \tilde{D}_{ki,\mathrm{PDR}}$. 

According to (\ref{qi}), we have that
\begin{eqnarray}
\lefteqn{\hat{Q}_{i,*} (\hat{\mu}_{*},\hat{\vec{\xi}})} \nonumber \\
&=& \left\{\frac{1}{(n-1)(n-2)}
\sum_{i=1}^{n}\sum_{\stackrel{\ell =i}{\ell  \ne i}}^{n}\sum_{\stackrel{r = 1}{r \ne \ell , r \ne i}}^{n} \frac{\partial G_{i\ell r,*}(\hat{\mu}_{*},\vec{\xi})}{\partial \vec{\xi}^\top}\bigg|_{\vec{\xi} = \hat{\vec{\xi}}}
\right\} \nonumber\\
&& \times \: \left\{-\sum_{i=1}^{n}\frac{\partial \mathcal{S}_i(\vec{\xi})}{\partial \vec{\xi}^\top}\bigg|_{\vec{\xi} = \hat{\vec{\xi}}}\right\}^{-1} \mathcal{S}_i(\hat{\vec{\xi}}) \nonumber \\
&& + \: \frac{1}{2(n-1)(n-2)} \sum_{\stackrel{\ell =1}{\ell  \ne i}}^{n} \sum_{\stackrel{r = 1}{r \ne i, r \ne \ell }}^{n}\bigg\{ G_{i\ell r,*}(\hat{\mu}_{*},\hat{\vec{\xi}}) + G_{ir \ell ,*}(\hat{\mu}_{*},\hat{\vec{\xi}}) + G_{\ell ir,*}(\hat{\mu}_{*},\hat{\vec{\xi}}) \nonumber\\
&& + \: G_{\ell r i,*}(\hat{\mu}_{*},\hat{\vec{\xi}}) + G_{r i\ell ,*}(\hat{\mu}_{*},\hat{\vec{\xi}}) + G_{r \ell i,*}(\hat{\mu}_{*},\hat{\vec{\xi}})\bigg\}. \nonumber
\end{eqnarray}
In addition, for fixed $i$, we also have that
\begin{eqnarray}
\sum_{\stackrel{\ell  = 1}{\ell  \ne i}}^{n} \sum_{\stackrel{r = 1}{r \ne i, r \ne \ell }}^{n} \left\{G_{i\ell r,*}(\hat{\mu}_{*},\hat{\vec{\xi}}) + G_{ikr,*}(\hat{\mu}_{*},\hat{\vec{\xi}})\right\} &=& 2\sum_{\stackrel{\ell  = 1}{\ell  \ne i}}^{n} \sum_{\stackrel{r = 1}{r \ne i, r \ne \ell }}^{n}G_{i\ell r,*}(\hat{\mu}_{*},\hat{\vec{\xi}}), \nonumber \\
\sum_{\stackrel{\ell  = 1}{\ell  \ne i}}^{n} \sum_{\stackrel{r = 1}{r \ne i, r \ne \ell }}^{n} \left\{G_{\ell ir,*}(\hat{\mu}_{*},\hat{\vec{\xi}}) + G_{r i\ell ,*}(\hat{\mu}_{*},\hat{\vec{\xi}})\right\} &=& 2\sum_{\stackrel{\ell  = 1}{\ell  \ne i}}^{n} \sum_{\stackrel{r = 1}{r \ne i, r \ne \ell }}^{n}G_{\ell ir,*}(\hat{\mu}_{*},\hat{\vec{\xi}}), \nonumber \\
\sum_{\stackrel{\ell  = 1}{\ell  \ne i}}^{n} \sum_{\stackrel{r = 1}{r \ne i, r \ne \ell }}^{n} \left\{G_{\ell r i,*}(\hat{\mu}_{*},\hat{\vec{\xi}}) + G_{r \ell i,*}(\hat{\mu}_{*},\hat{\vec{\xi}})\right\} &=& 2\sum_{\stackrel{\ell =1}{\ell  \ne i}}^{n} \sum_{\stackrel{r = 1}{r \ne i, r \ne \ell }}^{n}G_{r \ell i,*}(\hat{\mu}_{*},\hat{\vec{\xi}}). \nonumber
\end{eqnarray}
Therefore, 
\begin{eqnarray}
\lefteqn{\hat{Q}_{i,*} (\hat{\mu}_{*},\hat{\vec{\xi}})} \nonumber \\
&=& \left\{\frac{1}{(n-1)(n-2)}\sum_{i = 1}^{n}\sum_{\stackrel{\ell  = i}{\ell  \ne i}}^{n}\sum_{\stackrel{r = 1}{r \ne \ell , r \ne i}}^{n} \frac{\partial G_{i\ell r,*}(\hat{\mu}_{*},\vec{\xi})}{\partial \vec{\xi}^\top}\bigg|_{\vec{\xi} = \hat{\vec{\xi}}}\right\} \nonumber\\
&& \times \: \left\{-\sum_{i=1}^{n}\frac{\partial \mathcal{S}_i(\vec{\xi})}{\partial \vec{\xi}^\top}\bigg|_{\vec{\xi} = \hat{\vec{\xi}}}\right\}^{-1} \mathcal{S}_i(\hat{\vec{\xi}}) \label{asy:var:Qi} \\
&& + \: \frac{1}{(n-1)(n-2)} \sum_{\stackrel{\ell  = 1}{\ell  \ne i}}^{n} \sum_{\stackrel{r = 1}{r \ne i, r \ne \ell }}^{n}\bigg\{ G_{i\ell r,*}(\hat{\mu}_{*},\hat{\vec{\xi}}) +  G_{\ell ir,*}(\hat{\mu}_{*},\hat{\vec{\xi}}) + G_{r \ell i,*}(\hat{\mu}_{*},\hat{\vec{\xi}})\bigg\}. \nonumber
\end{eqnarray}
The quantity $\sum \limits_{i=1}^{n} \dfrac{\partial \mathcal{S}_i(\vec{\xi})}{\partial \vec{\xi}^\top}\bigg|_{\vec{\xi} = \hat{\vec{\xi}}}$ could be obtained as the Hessian matrix of the log-likelihood function at $\hat{\vec{\xi}}$. In order to compute $\dfrac{\partial G_{i\ell r,*}(\hat{\mu}_{*},\vec{\xi})}{\partial \vec{\xi}^\top}\bigg|_{\vec{\xi} = \hat{\vec{\xi}}}$, we have to get the derivatives $\dfrac{\partial}{\partial \vec{\xi}^\top} \rho_{ki}(\vec{\tau}_{0\rho_k})$, $\dfrac{\partial}{\partial \vec{\xi}^\top} \rho_{k(0)i}(\vec{\xi})$, $\dfrac{\partial}{\partial \vec{\xi}^\top} \pi^{-1}_{i}(\vec{\lambda}, \vec{\tau}_\pi)$, $\dfrac{\partial}{\partial \vec{\xi}^\top} \pi_{10i}(\vec{\lambda}, \vec{\tau}_\pi)$, $\dfrac{\partial}{\partial \vec{\xi}^\top} \pi_{01i}(\vec{\lambda}, \vec{\tau}_\pi)$ and $\frac{\partial}{\partial \vec{\xi}^\top} \pi_{00i}(\vec{\lambda}, \vec{\tau}_\pi)$.

In Section \ref{sec:para_est}, we obtain 
\begin{equation}
\begin{array}{r l r l}
\dfrac{\partial}{\partial \lambda_1} \pi_{10i}(\vec{\lambda}, \vec{\tau}_\pi) &= \pi_{10i}(1 - \pi_{10i}); &  \dfrac{\partial}{\partial \lambda_2} \pi_{10i}(\vec{\lambda}, \vec{\tau}_\pi) &= 0; \\ [16pt]
\dfrac{\partial}{\partial \lambda_1} \pi_{01i}(\vec{\lambda}, \vec{\tau}_\pi) &= 0; &  \dfrac{\partial}{\partial \lambda_2} \pi_{01i}(\vec{\lambda}, \vec{\tau}_\pi) &= \pi_{01i}(1 - \pi_{01i});  \\ [16pt]
\dfrac{\partial}{\partial \lambda_1} \pi_{00i}(\vec{\lambda}, \vec{\tau}_\pi) &= 0; &  \dfrac{\partial}{\partial \lambda_2} \pi_{00i}(\vec{\lambda}, \vec{\tau}_\pi) &= 0.
\end{array}
\nonumber
\end{equation}
and 
\[
\frac{\partial}{\partial \vec{\tau}_\pi^\top}\pi_{d_1 d_2 i} = \vec{U}_i (1 - \pi_{d_1 d_2 i})\pi_{d_1 d_2 i},
\]
where $(d_1, d_2)$ belongs to the set $\{(1,0), (0,1), (0,0)\}$.  Also, we have 
\begin{equation}
\begin{array}{r l r l}
\dfrac{\partial}{\partial \vec{\tau}^\top_{\rho_1}} \rho_{1i}(\tau_\rho) &= \vec{U}_i\rho_{1i}(1 - \rho_{1i}); &  \dfrac{\partial}{\partial \vec{\tau}^\top_{\rho_2}} \rho_{1i}(\vec{\tau}_\rho) &= - \vec{U}_i\rho_{1i}\rho_{2i}; \\ [16pt]
\dfrac{\partial}{\partial \vec{\tau}^\top_{\rho_2}} \rho_{2i}(\vec{\tau}_\rho) &= \vec{U}_i\rho_{2i}(1 - \rho_{2i}); &  \dfrac{\partial}{\partial \vec{\tau}^\top_{\rho_1}} \rho_{2i}(\vec{\tau}_\rho) &= - \vec{U}_i \rho_{1i}\rho_{2i}.
\end{array}
\nonumber
\end{equation}
Moreover,
\[
\frac{\partial}{\partial \lambda_s} \pi^{-1}_i(\vec{\lambda}, \vec{\tau}_\pi) = -D_{si}\frac{1 - \pi_i}{\pi_i}; \qquad \frac{\partial}{\partial \vec{\tau}_\pi^\top}\pi^{-1}_{i}(\vec{\lambda}, \vec{\tau}_\pi) = -\vec{U}_{i}\frac{1 - \pi_i}{\pi_i},
\]
with $s = 1, 2$. Then, recall that
\begin{eqnarray}
\rho_{1(0)i} &=& \frac{(1 - \pi_{10i})\rho_{1i}}{(1 - \pi_{10i})\rho_{1i} + (1 - \pi_{01i})\rho_{2i} + (1 - \pi_{00i})\rho_{3i}}, \nonumber \\
\rho_{2(0)i} &=& \frac{(1 - \pi_{01i})\rho_{2i}}{(1 - \pi_{10i})\rho_{1i} + (1 - \pi_{01i})\rho_{2i} + (1 - \pi_{00i})\rho_{3i}}, \nonumber \\
\rho_{3(0)i} &=& \frac{(1 - \pi_{00i})\rho_{3i}}{(1 - \pi_{10i})\rho_{1i} + (1 - \pi_{01i})\rho_{2i} + (1 - \pi_{00i})\rho_{3i}}. \nonumber
\end{eqnarray}
After some algebra, we get
\begin{eqnarray}
\frac{\partial}{\partial \lambda_1} \rho_{1(0)i}(\vec{\xi}) &=& \frac{1}{z^2}\left[-\pi_{10i}(1 - \pi_{10i})\rho_{1i}\left\{(1 - \pi_{01i})\rho_{2i} + (1 - \pi_{00i})\rho_{3i} \right\} \right] \nonumber , \\
\frac{\partial}{\partial \lambda_2} \rho_{1(0)i}(\vec{\xi}) &=& \frac{1}{z^2} \rho_{1i}\rho_{2i}\pi_{01i}(1 - \pi_{01i}) (1 - \pi_{10i}) \nonumber , \\
\frac{\partial}{\partial \vec{\tau}_\pi^\top} \rho_{1(0)i}(\vec{\xi}) &=& -\frac{\vec{U}_i}{z^2} \rho_{1i}(1 - \pi_{10i}) \bigg\{ \rho_{2i}(1 - \pi_{01i})(\pi_{10i} - \pi_{01i}) \nonumber\\
&& + \: \rho_{3i}(1 - \pi_{00i})(\pi_{10i} - \pi_{00i})\bigg\} \nonumber , \\
\frac{\partial}{\partial \vec{\tau}_{\rho_1}^\top} \rho_{1(0)i}(\vec{\xi}) &=& \frac{\vec{U}_i}{z^2} \rho_{1i} (1 - \pi_{10i}) \left\{ \rho_{2i}(1 - \pi_{01i}) + \rho_{3i}(1 - \pi_{00i}) \right\} \nonumber , \\
\frac{\partial}{\partial \vec{\tau}_{\rho_2}^\top} \rho_{1(0)i}(\vec{\xi}) &=& -\frac{\vec{U}_i}{z^2} \rho_{1i}\rho_{2i} (1 - \pi_{10i}) (1 - \pi_{01i}) \nonumber.
\end{eqnarray}
Finally, we set $z = (1 - \pi_{10i})\rho_{1i} + (1 - \pi_{01i})\rho_{2i} + (1 - \pi_{00i})\rho_{3i}$, and get
\begin{eqnarray}
\frac{\partial}{\partial \lambda_1} \rho_{2(0)i}(\vec{\xi}) &=& \frac{1}{z^2} \rho_{1i}\rho_{2i}\pi_{10i}(1 - \pi_{10i}) (1 - \pi_{01i}) \nonumber , \\
\frac{\partial}{\partial \lambda_2} \rho_{2(0)i}(\vec{\xi}) &=& \frac{1}{z^2} \left[-\pi_{01i}(1 - \pi_{01i})\rho_{2i}\left\{(1 - \pi_{10i})\rho_{1i} + (1 - \pi_{00i})\rho_{3i} \right\} \right] \nonumber , \\
\frac{\partial}{\partial \vec{\tau}_\pi^\top} \rho_{2(0)i}(\vec{\xi}) &=& -\frac{\vec{U}_i}{z^2} \rho_{2i}(1 - \pi_{01i}) \bigg\{ \rho_{1i}(1 - \pi_{10i})(\pi_{01i} - \pi_{10i}) \nonumber \\
&& + \: \rho_{3i}(1 - \pi_{00i})(\pi_{01i} - \pi_{00i})\bigg\} \nonumber , \\
\frac{\partial}{\partial \vec{\tau}_{\rho_1}^\top} \rho_{2(0)i}(\vec{\xi}) &=& -\frac{\vec{U}_i}{z^2} \rho_{1i}\rho_{2i} (1 - \pi_{10i}) (1 - \pi_{01i}) \nonumber , \\
\frac{\partial}{\partial \vec{\tau}_{\rho_2}^\top} \rho_{2(0)i}(\vec{\xi}) &=& \frac{\vec{U}_i}{z^2} \rho_{2i} (1 - \pi_{01i}) \left\{ \rho_{1i}(1 - \pi_{10i}) + \rho_{3i}(1 - \pi_{00i}) \right\} \nonumber.
\end{eqnarray}
The derivative $\dfrac{\partial}{\partial \vec{\xi}^\top} \rho_{3(0)i}(\vec{\xi})$ can be computed by using the fact that $\rho_{3(0)i} = 1 - \rho_{1(0)i} - \rho_{2(0)i}$.

\section*{Appendix 2}
Here, we show that the estimating functions $G_{i\ell  r,*}$ are unbiased under the working disease and verification models. Recall that $\vec{\xi} = (\vec{\lambda}^\top, \vec{\tau}^\top_\pi, \vec{\tau}^\top_\rho)^\top$.
\begin{itemize}
	\item FI estimator. We have
	\begin{eqnarray}
	\E\left\{G_{i\ell  r,\mathrm{FI}}(\mu_0, \vec{\xi}_0)\right\} &=& \E \left\{ \rho_{1i}(\vec{\tau}_{0\rho}) \rho_{2\ell }(\vec{\tau}_{0\rho}) \rho_{3r}(\vec{\tau}_{0\rho}) (I_{i\ell r} - \mu) \right\} \nonumber \\
	&=& \E \left\{ \rho_{1i}\rho_{2\ell }\rho_{3r}(I_{i\ell  r} - \mu_0) \right\}. \nonumber
	\end{eqnarray}
	Hence, $\E\left\{G_{i\ell  r,\mathrm{FI}}(\mu_0, \vec{\xi}_0)\right\} = 0$ from (\ref{org:vus2}).
	\item MSI estimator. Consider $\E\left\{D_{ki,\mathrm{MSI}}(\vec{\xi}_0)|T_i, \vec{A}_i\right\}$.
	We have
	\begin{eqnarray}
	\E \left\{D_{ki,\mathrm{MSI}}(\vec{\xi}_0)|T_i, \vec{A}_i\right\} &=& \E\left\{V_i D_{ki} + (1 - V_i)\rho_{k(0)i}(\vec{\xi}_0)|T_i, \vec{A}_i\right\} \nonumber \\
	&=& \E \left[ \E\left\{V_i D_{ki} + (1 - V_i)\rho_{k(0)i}(\vec{\xi}_0)|T_i, \vec{A}_i, V_i \right\} | T_i, \vec{A}_i \right] \nonumber \\
	&=& \Pro(V_i = 1|T_i, \vec{A}_i)\E \left(D_{ki}|V_i = 1, T_i, \vec{A}_i\right) \nonumber\\
	&& + \: \Pro(V_i = 0|T_i, \vec{A}_i)\E \left(\rho_{k(0)i}(\vec{\xi}_0)|V_i = 0, T_i, \vec{A}_i \right) \nonumber \\
	&=& \Pro(V_i = 1|T_i, \vec{A}_i)\Pro(D_{ki} = 1|V_i = 1, T_i, \vec{A}_i) \nonumber\\
	&& + \: \Pro(V_i = 0|T_i, \vec{A}_i)\Pro(D_{ki} = 1|V_i = 0, T_i, \vec{A}_i) \nonumber\\
	&=& \Pro(D_{ki} = 1|T_i, \vec{A}_i) = \rho_{ki} \nonumber.
	\end{eqnarray}
	Therefore, 
	\begin{eqnarray}
	\E\left\{G_{i\ell r, \mathrm{MSI}}(\mu_0,\vec{\xi}_0) \right\} &=& \E \left\{D_{1i,\mathrm{MSI}}(\vec{\xi}_0) D_{2\ell, \mathrm{MSI}}(\vec{\xi}_0) D_{3r, \mathrm{MSI}}(\vec{\xi}_0) \left(I_{i\ell r} - \mu_0 \right)\right\}
	\nonumber \\
	&=& \E \Big[ \left(I_{i\ell r} - \mu_0 \right) \E \left\{ D_{1i,\mathrm{MSI}}(\vec{\xi}_0) | T_i, \vec{A}_i \right\} \E \left\{ D_{2\ell,\mathrm{MSI}}(\vec{\xi}_0) | T_\ell, \vec{A}_\ell \right\} \nonumber \\
	&& \times \: \E \left\{ D_{3r,\mathrm{MSI}}(\vec{\xi}_0) | T_r, \vec{A}_r \right\} \Big]\nonumber \\
	&=& \E \left\{ \rho_{1i}\rho_{2\ell }\rho_{3r}(I_{i\ell r} - \mu_0) \right\}. \nonumber
	\end{eqnarray}
	\item IPW estimator. In this case, 
	\begin{eqnarray}
	\E \left(V_i D_{ki} \pi^{-1}_i(\vec{\xi}_0)|T_i, \vec{A}_i \right) &=& \pi_i^{-1}(\vec{\xi}_0) \E \left(V_i D_{ki}|T_i, \vec{A}_i\right) \nonumber\\
	&=& \pi^{-1}_i(\vec{\xi}_0) \E \left\{D_{ki} \E \left(V_i |D_{1i}, D_{2i}, T_i, \vec{A}_i\right) \big| T_i, \vec{A}_i\right\} \nonumber \\
	&=& \pi_i^{-1} \E \left(\pi_i D_{ki}|T_i, \vec{A}_i\right) = \rho_{ki}. \nonumber
	\end{eqnarray}
	Thus,
	\begin{eqnarray}
	\E\left\{G_{i\ell r, \mathrm{IPW}}(\mu_0, \vec{\xi}_0)\right\} &=& \E \left\{ \frac{V_i V_\ell V_r D_{1i} D_{2\ell} D_{3r}} {\pi_i(\vec{\xi}_0) \pi_\ell(\vec{\xi}_0) \pi_k(\vec{\xi}_0)} \left(I_{i\ell r} - \mu_0\right)\right\}
	\nonumber \\
	&=& \E \Big\{ \left(I_{i\ell r} - \mu_0\right) \E (V_i D_{1i} \pi^{-1}_i(\vec{\xi}_0)|T_i, \vec{A}_i) \E (V_\ell D_{2\ell} \pi^{-1}_\ell(\vec{\xi}_0)|T_\ell, \vec{A}_\ell) \nonumber \\
	&& \times \: \E (V_r D_{3r} \pi^{-1}_r(\vec{\xi}_0)| T_r, \vec{A}_r) \Big\} \nonumber \\
	&=& \E \left\{\rho_{1i} \rho_{2\ell } \rho_{3r}(I_{i\ell r} - \mu_0) \right\}. \nonumber
	\end{eqnarray}
	\item PDR estimator.
	\begin{eqnarray}
	\lefteqn{\E\left\{D_{ki, \mathrm{PDR}}(\vec{\xi}_0)|T_i, \vec{A}_i\right\}} \nonumber\\
	&=& \E \left[\E \left\{\frac{V_i D_{ki}}{\pi_i(\vec{\xi}_0)} - \rho_{k(0)i}(\vec{\xi}_0)\left(\frac{V_i}{\pi_i(\vec{\xi}_0)} - 1\right)\bigg| D_{1i}, D_{2i}, T_i, \vec{A}_i\right\} \bigg| T_i, \vec{A}_i\right] \nonumber \\
	&=& \E \Bigg\{D_{ki} \E \left(\frac{V_i}{\pi_i(\vec{\xi}_0)} \bigg | D_{1i}, D_{2i}, T_i, \vec{A}_i\right) \nonumber \\
	&& - \: \rho_{k(0)i}(\vec{\xi}_0) \E \left(\frac{V_i}{\pi_i(\vec{\xi}_0)} - 1 \bigg | D_{1i}, D_{2i}, T_i, \vec{A}_i\right) \bigg| T_i, \vec{A}_i \Bigg\}
	\nonumber \\
	&=& \E(D_{ki} | T_i, \vec{A}_i) = \rho_{ki} \nonumber.
	\end{eqnarray}
	Hence,
	\begin{eqnarray}
	\E\left\{G_{i\ell r, \mathrm{PDR}}(\mu_0,\vec{\xi}_0)\right\} &=& \E \left\{D_{1i,\mathrm{PDR}}(\vec{\xi}_0) D_{2\ell, \mathrm{PDR}}(\vec{\xi}_0) D_{3r, \mathrm{PDR}}(\vec{\xi}_0) \left(I_{i\ell r} - \mu_0 \right)\right\}
	\nonumber \\
	&=& \E \Big[ \left(I_{i\ell r} - \mu_0 \right) \E \left\{ D_{1i,\mathrm{PDR}}(\vec{\xi}_0) | T_i, \vec{A}_i \right\} \E \left\{ D_{2\ell,\mathrm{PDR}}(\vec{\xi}_0) | T_\ell, \vec{A}_\ell \right\} \nonumber \\
	&& \times \: \E \left\{ D_{3r,\mathrm{PDR}}(\vec{\xi}_0) | T_r, \vec{A}_r \right\} \Big]\nonumber \\
	&=& \E \left\{ \rho_{1i}\rho_{2\ell }\rho_{3r}(I_{i\ell r} - \mu_0) \right\}. \nonumber
	\end{eqnarray}
\end{itemize}
 
\end{document}